\documentclass[a4paper,11pt]{article}
\pdfoutput=1
\usepackage{jheppubnew}
\usepackage{graphicx}
\usepackage{bm}
\usepackage{caption, subcaption}
\usepackage[all]{xy}
\usepackage{xcolor}
\usepackage{comment}

\newcommand{\be}{\begin{equation}}
\newcommand{\ee}{\end{equation}}

\newcommand{\D}{\mathrm{d}}
\newcommand{\C}{\mathbb{C}}
\newcommand{\R}{\mathbb{R}}
\newcommand{\Z}{\mathbb{Z}}

\renewcommand{\O}{\mathcal{O}}

\newcommand{\ep}{\epsilon}
\renewcommand{\a}{\alpha}

\newcommand{\nn}{\nonumber}
\newcommand{\lla}{\langle \! \langle}
\newcommand{\rra}{\rangle \! \rangle}
\renewcommand{\a}{\alpha}
\renewcommand{\b}{\beta}

\DeclareMathOperator{\re}{Re}

\DeclareMathOperator{\Li}{Li}

\newcommand{\bs}[1]{\boldsymbol{#1}}
\definecolor{darknavy}{RGB}{0,0,150} 
\definecolor{darkgreen}{rgb}{0,0.42,0.24}
\definecolor{darkred}{RGB}{177,35,35}  
\newcommand{\red}[1]{{\color{darkred} #1}}
\newcommand{\navy}[1]{{\color{darknavy} #1}}
\newcommand{\green}[1]{{\color{darkgreen} #1}}

\DeclareMathOperator{\K}{K}
\DeclareMathOperator{\B}{B}

\newcommand{\reg}[1]{\tilde{#1}}

\title{Evaluation of conformal integrals}
\author[a]{Adam Bzowski,}
\author[b]{Paul McFadden}
\author[c]{and Kostas Skenderis.}
\affiliation[a]{Institute for Theoretical Physics, K.U.~Leuven, Belgium.}
\affiliation[b]{Theoretical Physics Group, Blackett Laboratory, Imperial College, London, UK.}
\affiliation[c]{STAG research centre and Mathematical Sciences, University of Southampton, UK.}
\emailAdd{adam.bzowski@fys.kuleuven.be}
\emailAdd{p.mcfadden@imperial.ac.uk}
\emailAdd{k.skenderis@soton.ac.uk}

\begin{document}

\abstract{
We present a comprehensive method for the evaluation of a vast class of integrals 
representing 3-point functions of conformal field theories in momentum space.
The method leads to analytic, closed-form expressions for all scalar and tensorial 3-point functions of operators with integer dimensions in any spacetime dimension. In particular, this encompasses all 3-point functions of the stress tensor, conserved currents and marginal scalar operators.
}

\maketitle

\section{Introduction}

Conformal invariance imposes strong constraints on the form of correlation functions in any field theory. In particular, 2- and 3-point functions of the stress tensor, conserved currents and scalar operators are uniquely fixed up to a few constants. The standard approach, developed over several decades \cite{Polyakov:1970xd,Osborn:1993cr,Costa:2011a,Giombi:2011rz}, proceeds in position space and leads to position-space expressions. Recently, however, a pressing need for closed-form {\it momentum-space} expressions for correlators has arisen in various applications including cosmology \cite{McFadden:2009fg,McFadden:2010na,McFadden:2010vh,Bzowski:2012ih,Antoniadis:2011ib,Maldacena:2011nz, Creminelli:2012ed,
Kehagias:2012pd,Kehagias:2012td,Coriano:2012hd,Mata:2012bx,Ghosh:2014kba,Arkani-Hamed:2015bza}, the analysis of general properties of CFTs \cite{Giannotti:2008cv, Armillis:2009sm, Armillis:2009pq,Maldacena:2011jn,
Coriano:2012wp}, and condensed matter physics \cite{Chowdhury:2012km, Huh:2014eea}.

In the papers \cite{Bzowski:2013sza,Scalars} we initiated a comprehensive study of momentum-space 3-point functions in any conformal field theory.  
In  \cite{Bzowski:2013sza}, we expressed all 3-point functions involving the stress tensor, conserved currents and scalar operators of arbitrary dimension in terms of a class of scalar integrals we call \emph{triple-$K$ integrals}.
For special combinations of the operator and spacetime dimensions, these triple-$K$ integrals contain divergences necessitating their regularisation and renormalisation.  (For this same reason one cannot generally obtain the momentum-space correlators via a straightforward Fourier transform.)
In \cite{Scalars}, we presented a complete classification of the divergences and their renormalisation for purely scalar 3-point functions,
and in \cite{Bzowski:2017poo, Bzowski:2018fql} we discuss the corresponding renormalisation procedure for tensorial 3-point functions.

Having expressed all 3-point functions in terms of triple-$K$ integrals, the next step is to calculate them. 
In this paper we present a complete method for systematically computing 
all the triple-$K$ integrals 
that arise in the evaluation of scalar and tensorial 
3-point functions of operators with {\it integer dimensions}.   This broad class includes many operators of physical interest such as the stress tensor, conserved currents, and marginal scalar operators.

In cases where the spacetime dimension is odd, all  triple-$K$ integrals can be evaluated trivially in terms of elementary functions.  We therefore focus our attention on cases where the spacetime dimension is even.  
For purely scalar correlators, the only non-trivial new cases to be analysed are those for which both the following conditions hold:

\begin{enumerate}
\item[(i)] The spacetime dimension $d \geq 2$ is an even integer and all conformal dimensions $\Delta_j \in \Z$ are integers satisfying $\Delta_j \geq d/2$.
\item[(ii)] The following triangle inequalities are all satisfied:
\begin{equation} \label{e:triangle}
\Delta_1 + \Delta_2 > \Delta_3, \qquad \Delta_2 + \Delta_3 > \Delta_1, \qquad \Delta_3 + \Delta_1 > \Delta_2.
\end{equation}
\end{enumerate}
The first condition arises through a special symmetry property of the triple-$K$ integral which allows us to relate cases with $\Delta_j<d/2$ to those with $\Delta_j>d/2$.  
As for the second condition, if any of the triangle inequalities are violated then we have no need to evaluate the triple-$K$ integral: in such cases the triple-$K$ integral diverges and the renormalised 3-point function is simply proportional to the leading divergence,  as described in \cite{Scalars}.\footnote{With integer $\Delta_j$, the violation of condition (ii) implies the presence of $(++-)$-type singularities in the 
terminology
of \cite{Scalars}.}  Crucially this leading divergence, and hence the 3-point function, can be extracted through a simple series expansion of the integrand without evaluating the full triple-$K$ integral.  Explicit expressions for the renormalised 3-point functions in all such cases may be found in appendix A of \cite{Scalars}.

As we will discuss in section \ref{sec:tripleKdiv}, conditions (i) and (ii) can be re-expressed in terms of the parameters appearing in the corresponding triple-$K$ integrals.  
This equivalent form of the conditions (corresponding to (a)--(c) on page \pageref{eq2.9}) is very useful, since it also parametrises the non-trivial triple-$K$ integrals arising for correlators with tensorial structure.
Thus, to complete the analysis of all 3-point functions -- both scalar and tensorial -- featuring operators of integer dimensions, we must solve all triple-$K$ integrals for which this latter set of conditions hold.
As we will show, all such triple-$K$ integrals can be reduced in a finite number of steps to a single \emph{master integral}.  This integral  can   be evaluated in closed form and contains only a single special function, the dilogarithm $\Li_2$.

The organisation of this paper is as follows.
First, in section \ref{sec:tripleKdiv}, we review the triple-$K$ integral and elaborate in greater detail on role of conditions (i) and (ii).
We also recall the singularities that can arise in triple-$K$ integrals, their regularisation and how to pass between different schemes.  (This material is largely complementary to that in \cite{Scalars} although here we take a somewhat more mathematical focus.)
Next, our reduction scheme expressing the relevant triple-$K$ integrals  in terms of the master integral is presented in section \ref{sec:eval}. The scheme is recursive in nature and is based on simple identities between  Bessel functions.  Since both scalar and tensorial 1-loop 3-point massless Feynman integrals can be re-expressed as triple-$K$ integrals, our procedure generalises and simplifies other momentum-space 
recursion schemes such as those presented in \cite{Davydychev:1992xr,Campbell:1996zw,Ferroglia:2002mz}.  
For purposes of illustration, we apply our reduction scheme to evaluate all the triple-$K$ integrals arising in 3-point functions of the stress tensor and conserved currents in four-dimensional CFTs.
The evaluation of the master integral is then undertaken in section \ref{sec:MI}.
Its solution relies on expressing the master integral in terms of hypergeometric and Appell functions. (Similar methods for representing momentum integrals have appeared elsewhere in the literature, see for example \cite{ CabralRosetti:1998sp, Davydychev:1999mq, 
Anastasiou:1999ui, Boos:1987bg, Davydychev:2005nf, CabralRosetti:2002qv, Kniehl:2011ym, Coriano:2013jba}.)

Overall, our focus will be on 
computing the triple-$K$ integrals 
rather than the 3-point functions themselves.  
This approach makes sense 
as triple-$K$ integrals are the simple building blocks from which the generally more-complicated  3-point functions are constructed.  After the relevant triple-$K$ integrals have been evaluated via our recursive procedure, the full scalar and/or tensorial 3-point functions can be reconstructed as described in \cite{Bzowski:2013sza, Scalars}.

\section{Triple-$K$ integrals and their divergences} \label{sec:tripleKdiv}

\subsection{Overview}
\label{sec:review}

A {\it triple-$K$ integral} is a function of three momentum magnitudes $p_1, p_2, p_3$, defined as
\begin{equation} \label{e:tripleK}
I_{\alpha \{\beta_1 \beta_2 \beta_3\}}(p_1, p_2, p_3) = p_1^{\beta_1} p_2^{\beta_2} p_3^{\beta_3} \int_0^\infty \D x \: x^{\alpha} K_{\beta_1}(p_1 x) K_{\beta_2}(p_2 x) K_{\beta_3}(p_3 x).
\end{equation}
Here $K_{\nu}(z)$ denotes a modified Bessel function of the second kind, or Bessel $K$ function for short. The constants $\alpha$ and $\beta_j$ are fixed numbers relating to the physical input. For example, in the case of scalar operators $\O_1, \O_2, \O_3$ of respective dimensions $\Delta_1, \Delta_2, \Delta_3$ in $d$-dimensional conformal field theory, the unique 3-point function in momentum space is
\begin{equation} \label{e:3pt}
\lla \O_1(\bs{p}_1) \O_2(\bs{p}_2) \O_3(\bs{p}_3) \rra = c_{123} I_{\alpha \{\beta_1 \beta_2 \beta_3\}}(p_1, p_2, p_3),
\end{equation}
with
\begin{equation} \label{e:dims_to_params}
\alpha = \frac{d}{2} - 1, \qquad \beta_j = \Delta_j - \frac{d}{2}, \qquad j = 1,2,3,
\end{equation}
where $c_{123}$ denotes an unspecified theory-dependent constant. 
Triple-$K$ integrals are thus indeed conformal integrals as per our title; they satisfy appropriate dilatation and special conformal Ward identities as discussed in \cite{Bzowski:2013sza, Scalars}.
For tensorial 3-point functions, triple-$K$ integrals with various different $\alpha$ and $\beta_j$ can arise for a given set of operator and spacetime dimensions $\Delta_j$ and $d$; see \cite{Bzowski:2013sza} for a full description.
In physical situations the momentum magnitudes obey the triangle inequalities $p_i + p_j \geq p_k$ for all $i,j,k = 1,2,3$ as a consequence of momentum conservation  $\sum_j\bs{p}_j = 0$.  For the purposes of this paper, however, it will be sufficient to assume 
the $p_j$ are simply real positive numbers.

As noted above, the triple-$K$ integral is related to massless 1-loop 3-point Feynman integrals in momentum space.  The exact relation, derived in appendix A.3 of \cite{Bzowski:2013sza}, is reproduced in appendix \ref{sec:momentum} for convenience.
Triple-$K$ integrals are also naturally related to   holographic 3-point functions, since the AdS bulk-to-boundary propagator in Poincar\'{e} coordinates is proportional to the Bessel $K$ function.  The representation \eqref{e:3pt} thus arises in holographic calculations of 3-point functions
\cite{Freedman:1998tz} (see also appendix D of \cite{Scalars}). 

As it stands, the triple-$K$ integral \eqref{e:tripleK} converges in the range
\begin{equation} \label{e:conv}
\alpha + 1 > | \beta_1 | + | \beta_2 | + | \beta_3 |,
\end{equation}
with fixed $p_1, p_2, p_3 > 0$.  Outside this range, the triple-$K$ integral can be defined through its unique analytic continuation. In fact, we will always regard the triple-$K$ integral as a maximally extended analytic function which, on its domain of convergence, agrees with \eqref{e:tripleK}.

The triple-$K$ integral defined in this manner still exhibits singularities at special values of $\alpha$ and $\beta_j$, as shown in \cite{Scalars}. These special values correspond to solutions of the condition 
\begin{equation} \label{e:cond}
\alpha + 1 \pm \beta_1 \pm \beta_2 \pm \beta_3 = - 2 n, \qquad n = 0,1,2,\ldots
\end{equation}
The triple-$K$ integral becomes singular if there exists any choice of independent signs and non-negative integer $n$ such that the above condition is satisfied.  It is often useful to refer to these singularities by their associated set of signs; thus, for example, if \eqref{e:cond} is satisfied with a $++-$ choice of signs then we will call the resulting singularity a $(++-)$ singularity. 
Note that there may exist more than one choice of signs for any specific value of $n$.

For physical applications, if the triple-$K$ integral diverges
we have to regulate and renormalise.
If the condition \eqref{e:cond} holds, we introduce the regulated parameters
\begin{equation} \label{e:reg}
\a \mapsto \reg{\a} = \a + u \ep, \qquad\qquad \b_j \mapsto \reg{\b}_j = \b_j + v_j \ep,
\end{equation}
where $u$ and $v_j$, $j=1,2,3$ are fixed but arbitrary numbers. In this way we regard the regulated triple-$K$ integral
\begin{equation} \label{e:ireg}
I_{\alpha \{ \beta_1 \beta_2 \beta_3\}}(p_1, p_2, p_3) \ \longmapsto \ I_{\reg{\alpha} \{ \reg{\beta}_1 \reg{\beta}_2 \reg{\beta}_3\}}(p_1, p_2, p_3)
\end{equation}
as a function of the regulator $\epsilon$ with all momenta fixed. The divergence of the triple-$K$ integral manifests itself as a pole at $\ep = 0$.
 Most of the integrals considered in this paper are divergent, meaning that such a regularisation is usually necessary.  
 
A full analysis of the singularity structure of triple-$K$ integrals and the associated renormalisation procedure for scalar 3-point functions was carried out in \cite{Scalars}.  In particular, the singular part of a triple-$K$ integral can always be extracted through a simple series expansion of its integrand,
meaning the singularities can be determined without a full evaluation of the integral.
In certain special cases, knowing these singularities alone is sufficient to determine the renormalised correlator.  More generally, however, to determine the renormalised 3-point function we also need to know the finite part of the regulated triple-$K$ integral as $\ep\rightarrow 0$. While the general method for obtaining this finite part has been sketched in \cite{Bzowski:2013sza}, our aim here is to present a more thorough analysis.

Let us return now to the conditions (i) and (ii) specified in the introduction.  Cases where all operator dimensions are integral but the spacetime dimension is {\it odd} are trivial since all the $\beta_j$ are half-integer.  (Recall \eqref{e:dims_to_params} for scalar correlators; for tensorial correlators the $\beta_j$ also turn out to be half-integer, see \cite{Bzowski:2013sza}.)
When all the $\beta_j$ are half-integer (as also occurs for operators of half-integral dimension in an even-dimensional spacetime), the Bessel $K$ functions in the integrand of the triple-$K$ integral \eqref{e:tripleK} reduce to elementary functions.
The entire triple-$K$ integral can then be evaluated in terms of elementary functions; the general result is listed in appendix \ref{sec:halfint}.

For operators of integer dimension in a {\it even}-dimensional spacetime, 
the $\beta_j$ are instead all integers and the Bessel $K$ functions are no longer elementary.
We can however restrict our attention to cases where all the $\beta_j\ge 0$. 
Since Bessel $K$ functions are even in their index, {\it i.e.,} $K_{\beta}(x)=K_{-\beta}(x)$, 
it follows immediately that
\begin{equation} \label{e:prered_scheme}
I_{\alpha \{ - \beta_1, \beta_2, \beta_3 \}} = p_1^{-2 \beta_1} I_{\alpha \{ \beta_1 \beta_2 \beta_3 \}},
\end{equation}
with similar identities holding for $\beta_2$ and $\beta_3$.  We can therefore relate any triple-$K$ integral in which some of the $\beta_j$ are negative to an equivalent triple-$K$ integral in which all $\beta_j\ge 0$.  For scalar correlators this is the reason we only needed to focus on cases where all the $\Delta_j\ge d/2$ as specified in condition (i).

Since our analysis of triple-$K$ integrals in the rest of the paper will be phrased in terms of the parameters $\alpha$ and $\beta_j$, it will be useful to re-state conditions (i) and (ii) directly in terms of these parameters.
This leads us to an equivalent set of conditions:
\begin{enumerate}
\item[(a)] All the $\beta_j$ are non-negative integers (i.e., $\beta_j \in 0, 1, 2\ldots$);
\item[(b)] The combination
\begin{equation}\label{eq2.9}
\alpha + 1 - \beta_1 - \beta_2 - \beta_3 = - 2 n_0,
\end{equation}
is an even integer, \textit{i.e.}, $n_0$ is an integer of any sign or zero;
\item[(c)] The following inequalities hold
\begin{equation}\label{eq2.10}
\alpha + 1 + \beta_1 + \beta_2 > \beta_3, \qquad \alpha + 1 + \beta_2 + \beta_3 >\beta_1, \qquad \alpha + 1  + \beta_1 + \beta_3>\beta_2.
\end{equation}
\end{enumerate}
Our reasons for writing the conditions in this precise form is partly for later convenience, as will become apparent.   Nevertheless, for scalar correlators with $\alpha$ and $\beta_j$ as in \eqref{e:dims_to_params}, note that given condition (a), condition (b) implies that the spacetime dimension is an even integer.  Condition (a) then yields that all $\Delta_j$ are integers satisfying $\Delta_j\ge d/2$, which is equivalent to condition (i).  Condition (c) is directly equivalent to condition (ii).  More generally, for tensor correlators where a range of $\alpha$ and $\beta_j$ parameters can appear for a given set of operator and spacetime dimensions, it will be more convenient simply to  use the conditions (a)--(c) in place of (i) and (ii).

The relation between the conditions (a)--(c) and the singularity condition \eqref{e:cond} for the triple-$K$ integral should also be noted.
Condition (c) forbids the appearance of $(++-)$, $(-++)$ and $(+-+)$ solutions of \eqref{e:cond}.  
Solutions of type $(+++)$ are also forbidden in view of condition (a).
Conversely, given conditions (a) and (b), violations of condition (c) implies that $(++-)$ singularities (and permutations) are necessarily present, and potentially also $(+++)$ singularities.
As mentioned in the introduction, if either $(+++)$ or $(++-)$ solutions (in any permutation) are present, the renormalised 3-point functions are simply given by the leading divergence of the triple-$K$ integral as $\ep\rightarrow 0$.  As shown in \cite{Scalars}, this leading divergence is nonlocal in the momenta and satisfies the complete set of homogeneous conformal Ward identities.
Thus, when condition (c) is violated, we have no need to evaluate the corresponding triple-$K$ integral: the leading singularity, and hence the renormalised 3-point function, can be extracted through a series expansion of the integrand.
For scalar 3-point functions a complete listing of the renormalised correlators is given in appendix A of \cite{Scalars}, while the equivalent analysis for tensorial correlators is given in  \cite{Bzowski:2017poo, Bzowski:2018fql}.

To re-iterate, our goal in this paper will be to evaluate all triple-$K$ integrals assuming the conditions (a)--(c) hold.
Our reduction scheme will enable all such integrals to be reduced to a single master integral of the form 
\begin{equation} \label{e:master1}
I_{\reg{0} \{\reg{1}\reg{1}\reg{1}\}} = I_{0 + u \epsilon \{ 1 + v_1 \epsilon, 1 + v_2 \epsilon, 1 + v_3 \epsilon\}},
\end{equation}
which can be evaluated in terms of elementary functions plus the dilogarithm.
Before presenting this reduction scheme, however, let us now review in greater detail the singularity structure of the regulated triple-$K$ integrals and their dependence on the choice of regularisation scheme.

\subsection{Analyticity in parameters}
\label{sec:analytic}

As discussed above, we wish to define the triple-$K$ integral through its maximal analytic continuation agreeing with \eqref{e:tripleK} on its domain of convergence.  We must therefore show that the integral is analytic with respect to the four parameters $\alpha$ and $\beta_j$ for $j=1,2,3$. 
Introducing a complexified space of parameters $(\alpha, \beta_1, \beta_2, \beta_3) \in \C^4$, we wish to show that the triple-$K$ integral, regarded as a function
\begin{equation} \label{e:complex_fun}
(\alpha, \beta_1, \beta_2, \beta_3) \longmapsto I_{\alpha \{ \beta_1 \beta_2 \beta_3\}}(p_1, p_2, p_3)
\end{equation}
with fixed positive $p_1, p_2, p_3 > 0$, is analytic in the region of convergence
\begin{equation} \label{e:domain}
\re (\alpha  - \beta_t + 1) > 0, \qquad \re \beta_j > 0, \qquad j = 1,2,3,
\end{equation}
where $\beta_t = \beta_1 + \beta_2 + \beta_3$.

By Hartogs' theorem a complex function of many variables is analytic if it is analytic in each variable separately. Analyticity in a single variable can be proven by means of Morera's theorem, \textit{i.e.}, by showing that an indefinite integral of the function \eqref{e:complex_fun} exists. For this one shows that the integral over any closed curve in the parameter space vanishes. 

For concreteness consider a closed curve $C$ in the complex plane of $\beta_3$. Since for $x > 0$ we have $|x^{\alpha}| \leq x^{|\alpha|}$ and $|K_{\beta}(x)| \leq 2 K_{|\beta|}(x)$, using Fubini's theorem we can write
\begin{align}
& \int_\gamma \D \beta_3 \int_0^\infty \D x \: x^{\alpha} p_1^{\beta_1} p_2^{\beta_2} p_3^{\beta_3} K_{\beta_1}(p_1 x) K_{\beta_2}(p_2 x) K_{\beta_3}(p_3 x) \nn\\
& \qquad = \int_0^\infty \D x \: x^{\alpha} p_1^{\beta_1} p_2^{\beta_2} K_{\beta_1}(p_1 x) K_{\beta_2}(p_2 x) \int_\gamma \D \beta_3 \: p_3^{\beta_3} K_{\beta_3}(p_3 x) \nn\\
& \qquad = 0
\end{align}
In the last line we used the analyticity of the function $\beta_3 \mapsto p_3^{\beta_3} K_{\beta_3}(p_3 x)$ and Cauchy's theorem.  The triple-$K$ integral is therefore indeed analytic with respect to $\beta_3$ and an identical argument applies to the remaining parameters.

Since the triple-$K$ integral is analytic with respect to all its parameters in its non-empty domain of convergence \eqref{e:domain}, the method of analytic continuation can be applied. Outside the domain of convergence \eqref{e:domain}, the unique analytically continued function may exhibit singularities; a possibility we would now like to explore.

\subsection{Structure of singularities}
\label{sec:when}

As discussed above, singularities in the triple-$K$ integral arise if the condition \eqref{e:cond},
\begin{equation} \label{e:cond1}
\alpha + 1 \pm \beta_1 \pm \beta_2 \pm \beta_3 = - 2 n
\end{equation}
is satisfied for at least one independent choice of signs and a non-negative integer $n$ \cite{Scalars}. 
The singular behaviour of the triple-$K$ integral arises from the divergence at its lower limit $x=0$. Indeed, as $K_{\nu}(x) \sim x^{-1/2} e^{-x}$ for large $x$, the integral always converges at $x = \infty$.

To determine the pole structure of the singularity, it suffices to expand the integrand about $x=0$ using \eqref{e:expK} and \eqref{e:expKn}. Since we regard expressions such as \eqref{e:tripleK} through their maximal analytic extensions, the integral diverges if there exists a term of order $1/x$ in the expansion of the integrand. Indeed, each power term $x^a$ integrates to
\begin{equation} \label{e:analcont1}
\int_0^{\mu^{-1}} x^a \D x = \frac{\mu^{-(a+1)}}{a + 1},
\end{equation}
where the upper limit of the integral is arbitrary. In particular, the divergent part of the integral cannot depend on $\mu$. While the convergence of the integral requires $a > -1$, the right-hand side of the above expression is an analytic function of $a \in \C$ with a single pole at $a=-1$. It therefore defines an analytic extension of the integral for any $a \in \C \backslash \{-1\}$. Thus, while the triple-$K$ integral naively diverges if the expansion of its integrand contains terms of the form $x^a$ with real $a < -1$, its value in such cases is in fact uniquely defined through the analytic continuation \eqref{e:analcont1}. The condition \eqref{e:cond} simply enumerates all possible instances where terms of the form $1/x$ appear in the expansion of the integrand.  The coefficients of the various poles then follow from the series expansions \eqref{e:expK} and \eqref{e:expKn} of the Bessel functions. 

If all the $\beta_j$ are non-integer, 
the series expansion of the Bessel functions just consists of powers $x^a$ for various exponents $a$.  If, however, some $\beta_j$ are integral, logarithms can appear in the series expansion of the integrand according to \eqref{e:expKn}. The integral
\begin{equation} \label{e:analcont}
\int_0^{\mu^{-1}} x^{a} \log^n x \: \D x = \frac{(-1)^n n!}{(a + 1)^{n+1}} \mu^{-(a + 1)} \sum_{j=0}^n \frac{(a+1)^j \log^j \mu}{j!},
\end{equation}
then generates a pole of order $n + 1$ at $a = -1$. As the function is analytic away from the singularity, the order of the pole does not depend on the `direction' of approach. In particular, the series expansion around $a = -1$ reads
\begin{equation} \label{e:analcontexp}
\int_0^{\mu^{-1}} x^{a} \log^n x \: \D x = \frac{(-1)^n n!}{(a + 1)^{n+1}} + \frac{(-1)^n \log^{n+1} (\mu^{-1})}{n + 1} + O(a+1).
\end{equation}
As we can see, the only divergent term is the leading pole of order $n+1$, and in agreement with our expectations, the scale $\mu$ is absent in this term.

\subsection{Regularisation scheme} \label{ch:regul_schemes}

For physical applications it is convenient to analyse divergent integrals by introducing a regulator. We start with a divergent triple-$K$ integral with fixed parameters $\alpha$ and $\beta_j$ and satisfying at least one of the conditions \eqref{e:cond1}. We then regulate the integral by shifting its parameters by small amounts proportional to a regulator $\epsilon$ according to the formula \eqref{e:ireg},
\begin{equation} \label{e:iireg}
I_{\alpha \{\beta_1 \beta_2 \beta_3\}} \ \longmapsto \ I_{\alpha + u \epsilon \{\beta_1 + v_1 \epsilon, \beta_2 + v_2 \epsilon, \beta_3 + v_3 \epsilon\}}.
\end{equation}
The fixed, but otherwise arbitrary numbers $u$, $v_1, v_2, v_3$ specify the direction of the shift. In general the regulated integral exists, but exhibits singularities when $\epsilon$ is taken to zero. As we will see, not all choices of directions parametrised by the constants $u$ and $v_j$ actually regulate the integral, but `good' choices exist and there are many of them.

In the triple-$K$ representation \eqref{e:3pt} for the 3-point function of scalar operators in a CFT, the relation between the parameters in the integral and the physical dimensions is given by \eqref{e:dims_to_params}. In this case, the shift in spacetime and conformal dimensions on going to the regularised theory are
\begin{equation}
d \mapsto d + 2 u \epsilon, \qquad\Delta_j \mapsto \Delta_j + (u + v_j) \epsilon, \qquad  j = 1,2,3.
\end{equation}
Certain regularisation schemes may be more useful than others (see the related discussion in \cite{Scalars}).  Some particularly useful choices are:
\begin{enumerate}
\item $u = v_1 = v_2 = v_3$.  In terms of the physical dimensions, this scheme corresponds to shifting
\begin{equation} \label{e:dimreg}
d \mapsto d + 2 u \epsilon, \qquad \Delta_j \mapsto \Delta_j + 2 u \epsilon, \qquad  j = 1,2,3.
\end{equation}
Its most important feature is the fact that dimensions of sources corresponding to CFT operators (namely $d-\Delta_j$) do not change. While useful from a physical perspective, this scheme suffers from the drawback that it does not regulate the triple-$K$ integral in all cases.
\item $v_j = 0$ for all $j=1,2,3$. This scheme preserves the $\beta_j$ and hence the indices of the Bessel functions in the triple-$K$ integral. It is therefore particularly useful in cases where the indices of Bessel functions are half-integral,
as discussed in appendix \ref{sec:halfint}.
\item $-u = v_1 = v_2 = v_3$. As we will see later, the first step in our evaluation of the master integral will take place  in this scheme. Many other triple-$K$ integrals with integer indices for the Bessel functions are also naturally evaluated in this scheme.
\end{enumerate}

\subsection{Divergences and scheme dependence} \label{sec:change_scheme}

In this section we want to answer two important questions regarding the regulated triple-$K$ integrals. Firstly, we want to find a simple method for extracting the terms divergent in $\epsilon$ for any triple-$K$ integral without actually evaluating the entire integral. Secondly, since a regulated triple-$K$ integral depends on parameters $u$ and $v_j$, we want to extract the dependence of a finite order $\epsilon^0$ part of the integral on these parameters.
Throughout the section we will therefore assume that the unregulated parameters $\alpha$ and $\beta_j$  satisfy the singularity condition \eqref{e:cond1} for at least one choice of signs and non-negative integer $n$.

These two problems are closely related as can be anticipated from physical reasoning. Indeed, in any local quantum field theory one would expect that the divergent terms in regulated correlation functions should be computable without knowledge of the entire correlator. Furthermore, such divergences should be removable by local counterterms and hence should be of a certain special form. Finally, one would expect scheme-dependent terms to be related to divergent terms, since both are related to the scheme dependence introduced by counterterms.

As an example, let us consider the master integral $I_{\reg{0} \{\reg{1}\reg{1}\reg{1}\}}$.  From the analysis of section \ref{sec:when}, the master integral exhibits a double pole in the regulator so we can write
\begin{equation}
I_{\reg{0} \{\reg{1}\reg{1}\reg{1}\}} = \frac{I^{(-2)}_{\reg{0} \{\reg{1}\reg{1}\reg{1}\}}}{\ep^2} + \frac{I^{(-1)}_{\reg{0} \{\reg{1}\reg{1}\reg{1}\}}}{\ep} + I^{(0)}_{\reg{0} \{\reg{1}\reg{1}\reg{1}\}} + O(\epsilon).
\end{equation}
The first problem requires finding a simple procedure to evaluate 
$I^{(-2)}_{\reg{0} \{\reg{1}\reg{1}\reg{1}\}}$ and $I^{(-1)}_{\reg{0} \{\reg{1}\reg{1}\reg{1}\}}$.  In principle all terms here, including the finite one, depend moreover on $u$ and $v_j$. For the second problem we want to extract these $u$ and $v_j$-dependent terms from the finite part $I^{(0)}_{\reg{0} \{\reg{1}\reg{1}\reg{1}\}}$. We will refer to such contributions as scheme-dependent terms.

This notion of scheme-dependent terms is not however very well defined, as one can always include in such terms any scheme-independent expression. Hence, what we are really looking for is a general procedure for changing the regularisation scheme.  Let us assume a triple-$K$ integral is evaluated in two different schemes with the corresponding values of the $u$ and $v_j$ parameters being $u, v_j$ and $\bar{u}, \bar{v}_j$. We then look for a simple procedure to evaluate the difference
\begin{equation} \label{e:change}
I_{\alpha + u \ep \{ \beta_j + v_j \ep \}} - I_{\alpha + \bar{u} \ep \{ \beta_j + \bar{v}_j \ep \}} = I_{\alpha \{ \beta_j \}}^{(\bar{u}, \bar{v}_j) \mapsto (u, v_j)}
\end{equation}
up to and including finite terms of order $\epsilon^0$. As we will see in section \ref{ch:integral}, the master integral can be evaluated explicitly in a specific regularisation scheme with  $-u = v_1 = v_2 = v_3$. Since ultimately we are interested in the master integral evaluated in an arbitrary scheme, however, we need a scheme-changing procedure expressed by means of \eqref{e:change}.

In \cite{Scalars}, we presented a procedure for changing scheme based on the use of differential operators reducing the degree of divergence of a triple-$K$ integral. In this manner any divergent triple-$K$ integral can be reduced to finite, scheme-independent integrals and the resulting expressions can be integrated back (with respect to the momenta) to find the most general form of the scheme-dependent terms.  The undetermined constants of integration that arise in this method can be fixed through the explicit computation of triple-$K$ integrals in the simplifying limit where some of the momenta become small.

In this paper we present an alternative procedure for changing the regularisation scheme. 
The method is rather simpler and faster as it does not require solving any differential equations. Moreover, only the integration of power functions is required, in contrast to small-momentum limits of triple-$K$ integrals as in \cite{Scalars}.
In the following two subsections we first review the method for extracting the divergences of triple-$K$ integrals, then proceed to the problem of changing the regularisation scheme.

\subsubsection{Extracting the divergences}
\label{sec:extractdiv}

As discussed in section \ref{sec:when}, all divergences of the triple-$K$ integral follow from the $x=0$ region of integration. 
To extract the divergences, then, we can expand all three Bessel $K$ functions appearing in the integrand of the regulated triple-$K$ integral and integrate the resulting expression from zero to some arbitrary cut-off $\mu^{-1}$. 
Using analytic continuation to define the integral, only terms for which the power of $x$ is close to $-1$ contribute to the divergence and one finds
\begin{align} \label{e:div}
I^{\text{(div)}}_{\reg{\alpha} \{ \reg{\beta}_j \}} & = \sum_{\substack{w \in \R \\ n \in \{0,1,\ldots\}}} \int_0^{\mu^{-1}} \D x \: x^{-1 + w \epsilon} \log^n x \: c_{-1 + w \epsilon, n}(p_1, p_2, p_3), \nn\\
& = \sum_{\substack{w \in \R \\ n \in \{0,1,\ldots\}}} \frac{(-1)^n n!}{(w \ep)^{n+1}} \mu^{-w \epsilon} \sum_{j=0}^n \frac{(w \epsilon \log \mu)^j}{j!}\, c_{-1 + w \epsilon, n}(p_1, p_2, p_3),
\end{align}
where $c_{-1 + w \epsilon, n}$ represents the coefficient of $x^{-1 + w \epsilon} \log^n x$ in the expansion of the integrand of the regulated triple-$K$ integral. There are only finitely many nonvanishing terms of this form.

The coefficients $c_{-1 + w \epsilon, n}$ can be read off from the standard series expansion of the Bessel $K$ functions, (\ref{e:expK}, \ref{e:expKn}). In general they may be singular at $\epsilon = 0$.  However, in the scheme where all $v_j = 0$, the $\beta_j$ parameters are not regularised meaning $\reg{\beta}_j = \beta_j$ is independent of $\epsilon$.  In this scheme then all $c_{-1 + w \epsilon, n}$ are finite as $\epsilon \rightarrow 0$, and all poles in the triple-$K$ integral follow from integrating the various logarithmic terms in \eqref{e:div}. 
From this perspective the $v_j=0$ scheme is rather convenient for practical calculations, as discussed further in appendix A of \cite{Scalars}.

If all $v_j \neq 0$, on the other hand, we can express the $c_{-1 + w \epsilon, n}$ in terms of the Bessel expansion coefficients \eqref{e:cf_ap},
\begin{equation}
a_j^{\sigma}(\beta) = \frac{(-1)^j \Gamma(- \sigma \beta - j)}{2^{\sigma \beta + 2 j + 1} j!}. \label{e:cf}
\end{equation}
In this case no logarithmic terms appear in the integrand of \eqref{e:div} and the expression can be simplified as follows. Consider all choices of signs $\sigma_1, \sigma_2, \sigma_3 \in \{ \pm 1 \}$ and all non-negative integers $n_1, n_2, n_3$ such that the condition \eqref{e:cond1} is satisfied with $n = \sum_j n_j$, \textit{i.e.},
\begin{equation} \label{e:cond3}
\alpha + 1 + \sum_{j=1}^3 ( \sigma_j \beta_j + 2 n_j ) = 0.
\end{equation}
As no logarithms are present, we use \eqref{e:div}  with $n=0$ leading to
\begin{equation} \label{e:div1}
I^{\text{(div)}}_{\reg{\alpha} \{ \reg{\beta}_j \}} = \sum_{\text{cond}} \frac{\mu^{-w \epsilon}}{w \epsilon} \prod_{j=1}^3 p_j^{(1 + \sigma_1) \reg{\beta}_j + 2 n_j} a_{n_j}^{\sigma_j}(\reg{\beta}_j),
\end{equation}
where the sum is taken over all $\sigma_j$ and $n_j$, $j=1,2,3$ satisfying the condition \eqref{e:cond3} and 
\begin{equation}
w = u + \sum_{j=1}^3 \sigma_j v_j. \label{e:w}
\end{equation}

\subsubsection{Changing the regularisation scheme}
\label{sec:schemechangedetails}

By construction, \eqref{e:div} encodes all divergent contributions to the regulated triple-$K$ integral, \textit{i.e.},
\begin{equation}
I_{\reg{\alpha} \{ \reg{\beta}_j \}} = I^{\text{(div)}}_{\reg{\alpha} \{ \reg{\beta}_j \}} + O(\epsilon^0).
\end{equation}
In fact,   
$I^{\text{(div)}}$  
also contains {\it all scheme-dependent contributions} to the triple-$K$ integral:
\begin{equation}\label{e:schemeindep}
\frac{\partial}{\partial u} \left( I_{\reg{\alpha} \{ \reg{\beta_j} \}} - I^{\text{(div)}}_{\reg{\alpha} \{ \reg{\beta_j} \}} \right) = \frac{\partial}{\partial v} \left( I_{\reg{\alpha} \{ \reg{\beta_j} \}} - I^{\text{(div)}}_{\reg{\alpha} \{ \reg{\beta_j} \}} \right) = O(\epsilon).
\end{equation}
The scheme-changing expression in \eqref{e:change} is then given by
\begin{equation} \label{e:tobe_change}
I_{\alpha \{ \beta_j \}}^{(\bar{u}, \bar{v}_j) \mapsto (u, v_j)} = I^{\text{(div)}}_{\alpha + u \ep \{ \beta_j + v_j \ep \}} - I^{\text{(div)}}_{\alpha + \bar{u} \ep \{ \beta_j + \bar{v}_j \ep \}} + O(\epsilon).
\end{equation}
This formula, derived in appendix \ref{sec:proof}, provides an effective way to change the  regularisation scheme. If a triple-$K$ integral is known in one regularisation scheme, then by adding the above expression one can find the value of the integral in any other scheme. 
Crucially \eqref{e:tobe_change} contains only a finite number of terms from the series expansion of the integrand, and hence is easily computed for any given triple-$K$ integral.


Let us illustrate this method of changing the regularisation scheme with a worked example. 
Our goal will be evaluate the triple-$K$ integral $I_{2\{111\}}$ regularised in a generic scheme. This example can also be found in \cite{Scalars}, 
although our present method is rather simpler as only elementary integrals of powers need to be evaluated.

The divergent part of the integral, regulated in a general scheme, reads
\begin{align} \label{e:scheme_change}
I^{\text{(div)}}_{\reg{2} \{ \reg{1}  \reg{1} \reg{1} \}} & = 2^{v_t \epsilon}\Gamma(1 + v_1 \epsilon) \Gamma(1 + v_2 \epsilon) \Gamma(1 + v_3 \epsilon) \int_0^{\mu^{-1}} \D x x^{-1 + (u - v_t) \epsilon} \nn\\
& = \frac{1}{(u - v_t) \epsilon} + \left[ \frac{v_t}{v_t - u} (\gamma_E - \log 2) - \log \mu \right] + O(\epsilon),
\end{align}
where $v_t=\sum_jv_j$.
As we expect, the scale $\mu$ only shows up in the finite part of this expression. Notice also that the coefficient of $\log \mu$ is independent of $u$ and $v_j$, and hence the scheme-changing term \eqref{e:tobe_change} does not depend on $\mu$. Using \eqref{e:scheme_change}, we can now immediately write down the required expression for the triple-$K$ integral in some scheme $(u,v_j)$ given its value in another scheme $(\bar{u},\bar{v}_j)$, namely
\begin{align}
& I^{\text{(div)}}_{2 + u \epsilon\{ 1 + v_1 \epsilon, 1 + v_2 \epsilon, 1 + v_3 \epsilon \}} = I^{\text{(div)}}_{2 + \bar{u} \epsilon\{ 1 + \bar{v}_1 \epsilon, 1 + \bar{v}_2 \epsilon, 1 + \bar{v}_3 \epsilon \}}  \nn\\[1ex]
& \qquad + \frac{1}{\epsilon} \left[ \frac{1}{(u - v_t)} - \frac{1}{(\bar{u} - \bar{v}_t)} \right] + (\gamma_E - \log 2) \left[ \frac{u}{v_t - u} - \frac{\bar{u}}{\bar{v}_t - \bar{u}} \right] + O(\epsilon).
\end{align}

\section{Reduction scheme} \label{sec:eval}

In this section we present the complete reduction scheme allowing for the evaluation of any integral satisfying conditions (a) - (c) from section \ref{sec:review}. The reduction scheme relies on our knowledge of a single master integral
\begin{equation}
I_{0 + u \ep \{ 1 + v_1 \ep, 1 + v_2 \ep, 1 + v_3 \ep\}},
\end{equation}
which we will return to evaluate in section \ref{sec:MI}.

All the integrals accessible through our reduction scheme exhibit either a single or a double pole in the regulator, or else possess a finite $\epsilon \rightarrow 0$ limit.  The master integral itself has a double pole in the regulator, as we will find in section \ref{sec:MI}.
In practice, it turns out to be convenient to introduce two auxiliary  `master' integrals: the  linearly divergent integral $I_{2\{111\}}$ and the finite integral $I_{1\{000\}}$. 
As a first step, integrals with single poles can then be related to $I_{2\{111\}}$ while finite integrals can be related to $I_{1\{000\}}$.  Both the latter integrals can then be related to the original master integral $I_{0\{111\}}$ in a second step.  (Integrals with double poles meanwhile reduce to $I_{0\{111\}}$ in one step.)

In the remainder of this section, then, our goal will be to reduce every triple-$K$ integral satisfying conditions (a) - (c) in section \ref{sec:review} to one of the three integrals above according to its degree of divergence. 
For ease of reference, we have gathered the key equations relating the different master integrals in the table below, along with their explicit expressions although we will not need these in the following.

\begin{center}
\begin{tabular}{|l|c|c|c|}
\hline
Integral & Order of divergence & Reduction to $I_{0\{111\}}$ & Explicit expression \\ \hline \hline
$I_{0\{111\}}$ & $2$ & -- & \eqref{e:masterap} \\ \hline
$I_{2\{111\}}$ & $1$ & \eqref{e:I2111from_master} & \eqref{e:I2111} \\ \hline
$I_{1\{000\}}$ & $0$ & \eqref{e:I1000from_master} & \eqref{e:I1000} \\ \hline
\end{tabular}
\end{center}

\subsection{Definitions and simplifications}

From the definition of the triple-$K$ integral \eqref{e:tripleK} it follows that for any permutation $\sigma$ of the set $\{1,2,3\}$,
\begin{equation}
I_{\alpha \{ \beta_{\sigma(1)} \beta_{\sigma(2)} \beta_{\sigma(3)} \}}(p_1, p_2, p_3) = I_{\alpha \{ \beta_1 \beta_2 \beta_3 \}}(p_{\sigma^{-1}(1)}, p_{\sigma^{-1}(2)}, p_{\sigma^{-1}(3)}). \label{e:0id} 
\end{equation}
Also, as noted earlier, since Bessel functions are even in their index, {\it i.e.,} $K_{-\nu}(x) = K_{\nu}(x)$, we have
\begin{equation}
I_{\alpha \{- \beta_1 \beta_2, \beta_3 \}} = p_1^{-2 \beta_1} I_{\alpha \{ \beta_1 \beta_2 \beta_3 \}}. \label{e:2id}
\end{equation}
By combining these two relations we can always order the $\beta_j$ parameters and assume
\begin{equation} \label{e:order_beta}
\beta_1 \geq \beta_2 \geq \beta_3 \geq 0.
\end{equation}
We will assume such an ordering from now on.

Consider a triple-$K$ integral satisfying conditions (a) - (c) from section \ref{sec:review}. As we will explain shortly, the degree of divergence of the regulated integral can be discerned from the values of three constants $n_0, n_1$ and $n_2$ defined as follows,
\begin{align}
2 n_0 = \beta_1 + \beta_2 + \beta_3 - \alpha - 1, \label{e:n0def}\\
2 n_1 = \beta_1 + \beta_2 - \beta_3 - \alpha - 1, \label{e:n1def}\\
2 n_2 = \beta_1 - \beta_2 - \beta_3 - \alpha - 1. \label{e:n2def}
\end{align}
For integer $\alpha, n_0, n_1, n_2$ with $n_2 < 0$, we will find the triple-$K$ integral possesses a simple representation in terms of elementary functions and dilogarithms and can be reduced to the master integral $I_{0\{111\}}$.

Let us first show however that this last condition is equivalent to conditions (a) - (c) presented in section \ref{sec:review}. Indeed, conditions (a) and (b) together imply that $\alpha$, $n_0$, $n_1 = n_0 - \beta_3$ and $n_2 = n_1 - \beta_2$ are all integers. Condition (c) is equivalent to $n_2 < 0$. (Indeed, notice that by assuming the ordering \eqref{e:order_beta} the first inequality of (c) is the strongest and equivalent to $n_2 < 0$.)
Conversely, by subtracting equations \eqref{e:n0def} and \eqref{e:n1def} as well as \eqref{e:n1def} and \eqref{e:n2def} one finds that $\beta_2$ and $\beta_3$ are integer if $n_0$, $n_1$ and $n_2$ are. Integer $\alpha$ then implies integer $\beta_1$ as well.

In the following subsections we will present a (non-unique) reduction scheme leading to analytic expressions for all integrals under considerations. We divide all triple-$K$ integrals into three classes, depending on the order of singularity. More precisely, we have the following cases:
\begin{itemize}
\item If $n_0 < 0$, the triple-$K$ integral $I_{\alpha \{\beta_j\}}$ is finite and expressible in terms of $I_{1\{000\}}$.
\item If $n_0 \geq 0$ and $n_1 < 0$ then the $(---)$ form of the condition \eqref{e:cond} is satisfied; the triple-$K$ integral $I_{\tilde{\alpha} \{\tilde{\beta}_j\}}$ has a single pole in the regulator $\epsilon$ and is completely expressible in terms of $I_{2\{111\}}$.
\item If $n_0 \geq 0$ and $n_1 \geq 0$ then both $(--+)$ and $(---)$ conditions in \eqref{e:cond} are satisfied; the triple-$K$ integral $I_{\tilde{\alpha} \{\tilde{\beta}_j\}}$ has a double pole in the regulator $\epsilon$ and is completely expressible in terms of the master integral $I_{0\{111\}}$.
\end{itemize}

Before we discuss these cases further, let us make a few remarks. The values of $n_0$ and $n_1$ can be either positive or negative, but the corresponding $(---)$ and/or $(--+)$ conditions are only satisfied when $n_0$ and/or $n_1$ are non-negative. We will always assume that $n_2 < 0$, which ensures that the $(-++)$ condition is never satisfied.

If the $(-++)$ condition were to be satisfied, the procedure we present is not sufficient for the evaluation of the finite part of the triple-$K$ integral.  Nevertheless, as explained in the introduction, in this case the renormalised 3-point function is simply proportional to the leading divergence of the triple-$K$ integral.
This leading divergence can be extracted as described in section \ref{sec:extractdiv}; see also section 4.3.4 and appendix A of \cite{Scalars} for exact results. 
Here, we will concentrate instead on cases when the finite part of a triple-$K$ integral is indeed necessary for the evaluation of the corresponding 3-point function.

Notice also that due to the ordering \eqref{e:order_beta}, we always have 
\begin{equation} \label{e:n_ineqs}
n_0 \geq n_1 \geq n_2.
\end{equation}
The first equality only occurs if $\beta_3 = 0$ and the second equality is only possible  if $\beta_2 = \beta_3 = 0$.
From \eqref{e:n_ineqs}, it follows that if the $(+--)$ condition is satisfied, the $(-+-)$ condition is too.  Similarly, if the $(-+-)$ condition is satisfied, so is the $(--+)$ condition. It is crucial here that $\alpha$ and all the $\beta_j$ are integers. In general other
conditions besides these may be satisfied, but in such cases our reduction scheme will not be applicable.

\subsection{Identities} \label{sec:der}

Let us now list the important identities between triple-$K$ integrals we will use for the development of a reduction scheme. Two relations we have already mentioned are
\begin{align}
I_{\alpha \{ \beta_{\sigma(1)} \beta_{\sigma(2)} \beta_{\sigma(3)} \}}(p_1, p_2, p_3) &= I_{\alpha \{ \beta_1 \beta_2 \beta_3 \}}(p_{\sigma^{-1}(1)}, p_{\sigma^{-1}(2)}, p_{\sigma^{-1}(3)}), \\
I_{\alpha \{ - \beta_1 \beta_2, \beta_3 \}} &= p_1^{-2 \beta_1} I_{\alpha \{ \beta_1 \beta_2 \beta_3 \}}.\label{new_rel}
\end{align}
Three further important relations involving derivatives are
\begin{align}
I_{\alpha \{ \beta_1 \beta_2 \beta_3 \}} &= - \frac{1}{p_1} \frac{\partial}{\partial p_1} I_{\alpha-1 \{ \beta_1+1, \beta_2 \beta_3 \}}, \label{e:1id} \\
I_{\alpha+1 \{\beta_1+1, \beta_2, \beta_3\}} &= \left( 2 \beta_1 - p_1 \frac{\partial}{\partial p_1} \right) I_{\alpha \{\beta_1 \beta_2 \beta_3\}}, \label{e:red_scheme2} \\
I_{\alpha + 2 \{ \beta_1 \beta_2 \beta_3 \}} & = \K_j I_{\alpha \{\beta_1 \beta_2 \beta_3 \}}, \label{e:red_scheme23}
\end{align}
where in the last equation the index $j$ takes values $j=1,2,3$ and $\K_j$ denotes the conformal Ward identity operator introduced in \cite{Bzowski:2013sza}
\begin{equation} \label{e:KOp}
\K_j = \K_{j, \beta_j} = \frac{\partial^2}{\partial p_j^2} - \frac{2 \beta_j - 1}{p_j} \frac{\partial}{\partial p_j}.
\end{equation}
This operator depends on a single parameter $\beta_j$ which in this paper will always be equal to the corresponding $\beta_j$ of the triple-$K$ integral upon which $\K_j$ acts.

Finally, we have the useful identities
\begin{align}
I_{\alpha-1 \{ \beta_1 \beta_2 \beta_3 \}} & = \frac{1}{\alpha - \beta_t} \left[ p_1^2 I_{\alpha \{\beta_1 - 1, \beta_2, \beta_3 \}} + p_2^2 I_{\alpha \{\beta_1, \beta_2 - 1, \beta_3 \}} + p_3^2 I_{\alpha \{\beta_1, \beta_2, \beta_3 - 1\}} \right], \label{e:red_scheme3} \\
I_{\alpha-1 \{ \beta_1 \beta_2 \beta_3 \}} & = \frac{1}{\alpha - \beta_t} \B_{\beta_1 \beta_2 \beta_3} I_{\alpha - 2 \{ \beta_1 - 1, \beta_2 - 1, \beta_3 - 1\}}, \label{e:redbeta}
\end{align}
where $\beta_t = \beta_1 + \beta_2 + \beta_3$ and for later convenience we have defined 
\begin{align} \label{e:Bbeta}
\B_{\beta_1 \beta_2 \beta_3} & = p_1^2 \left(2 (\beta_2-1) - p_2 \frac{\partial}{\partial p_2} \right) \left(2 (\beta_3-1) - p_3 \frac{\partial}{\partial p_3} \right) + \text{cyclic permutations.}
\end{align}

Notice that among these identities only \eqref{e:red_scheme3} decreases the value of $\alpha$. In the remaining identities the operators appearing on the right-hand sides act to increase the value of $\alpha$ while either increasing or decreasing $\beta_j$ by integer amounts. 
The action of these operators can be summarised as follows:

\begin{center}
\begin{tabular}{|c|c|c|}
\hline
change in $\alpha$ & change in $\beta_j$ & equation \\ \hline\hline
- & $\beta_j \mapsto - \beta_j$ &  \eqref{new_rel} \\ \hline
$\alpha \mapsto \alpha + 1$ & $\beta_j \mapsto \beta_j - 1$ & \eqref{e:1id} \\ \hline
$\alpha \mapsto \alpha + 1$ & $\beta_j \mapsto \beta_j + 1$ & \eqref{e:red_scheme2} \\ \hline
$\alpha \mapsto \alpha + 2$ & - & \eqref{e:red_scheme23} \\ \hline
$\alpha \mapsto \alpha - 1$ & $\beta_j \mapsto \beta_j + 1$ & \eqref{e:red_scheme3} \\ \hline
$\alpha \mapsto \alpha + 1$ & all three $\beta_j \mapsto \beta_j + 1$ & \eqref{e:redbeta} \\ \hline
\end{tabular}
\end{center}

In the following two subsections we will prove the identities listed above.

\subsubsection{Relations involving derivatives}

Relation \eqref{e:1id} follows directly from a simple property of the Bessel function,
\begin{equation}
\frac{\partial}{\partial p} \left[ p^\nu K_{\nu}(p x) \right] = - x p^\nu K_{\nu - 1}(p x).
\end{equation}
If, however, we first use equation \eqref{new_rel}, followed by \eqref{e:1id} then \eqref{new_rel} again, we derive the second identity \eqref{e:red_scheme2},
\begin{align}
I_{\alpha+1 \{\beta_1+1, \beta_2, \beta_3\}} &= - p_1^{2 \beta_1 + 1} \frac{\partial}{\partial p_1} \left[ p_1^{-2 \beta_1} I_{\alpha \{\beta_1 \beta_2 \beta_3\}} \right] \nn\\
&= \left( 2 \beta_1 - p_1 \frac{\partial}{\partial p_1} \right) I_{\alpha \{\beta_1 \beta_2 \beta_3\}},
\end{align}
Furthermore, we can apply this operation and its permutations repeatedly to obtain
\begin{equation}
I_{\alpha + k_t \{\beta_j + k_j \}} = (-1)^{k_t} \left[\prod_{j=1}^3 p_j^{2(\beta_j + k_j)} \left( \frac{1}{p_j} \frac{\partial}{\partial p_j} \right)^{k_j} \right] \left[ p_1^{-2 \beta_1} p_2^{-2 \beta_2} p_3^{-2 \beta_3} I_{\alpha \{\beta_j\}} \right], \label{e:red_scheme2a}
\end{equation}
where $k_t = k_1 + k_2 + k_3$ and $k_j$ are non-negative integers.

Both the reduction relations \eqref{e:1id} and \eqref{e:red_scheme2} obtained by differentiation happen to increase $\alpha$ by one, while changing the value of one of the $\beta_j$ by $\pm 1$. When they are combined together, they increase the value of $\alpha$ by two and the resulting expression is equal to \eqref{e:red_scheme23}.

\subsubsection{Relations between various integrals}

To derive relation \eqref{e:red_scheme3} consider the following integral
\begin{align}
& \int_0^\infty \D x \: \frac{\partial}{\partial x} \Big( x^{\alpha} \prod_{j=1}^3 p_j^{\beta_j} K_{\beta_j}(p_j x) \Big) = (\alpha - \beta_t) I_{\alpha - 1 \{ \beta_1 \beta_2 \beta_3 \}} \nn\\
& \qquad\qquad\qquad
- \left[ p_1^2 I_{\alpha \{ \beta_1 - 1, \beta_2, \beta_3 \}} + p_2^2 I_{\alpha \{ \beta_1, \beta_2 - 1, \beta_3 \}} + p_3^2 I_{\alpha \{ \beta_1, \beta_2, \beta_3 - 1\}} \right].
\end{align}
In the domain of convergence \eqref{e:conv} the boundary term vanishes by simple power counting. Hence, by analytic continuation, relation \eqref{e:red_scheme3} holds whenever its two sides remain finite.

The second equation \eqref{e:redbeta} follows from a combination of \eqref{e:red_scheme3} with \eqref{e:red_scheme2} applied twice to each integral on the right-hand side.

Relation \eqref{e:red_scheme3} is the only relation reducing the value of $\alpha$. It is closely related to Davydychev's recursion relation (3.4) introduced in \cite{Davydychev:1992xr} as well as equation (36) of \cite{Davydychev:1995mq}. Indeed, using equation (A.3.17) of \cite{Bzowski:2013sza} one can rewrite Davydychev's $J$ integral defined in (2.1) of \cite{Davydychev:1992xr} as
\begin{equation}
J(\delta_1, \delta_2, \delta_3) = \frac{4 \pi^2}{\Gamma(\delta_1) \Gamma(\delta_2) \Gamma(\delta_3) \Gamma(4 - \delta_t)} I_{1 \{2-\delta_2-\delta_3, 2-\delta_1-\delta_3, 2-\delta_1-\delta_2\}}. \label{e:Jdav}
\end{equation}
Note that the rather complicated form of equation (3.4) in \cite{Davydychev:1992xr} is a consequence of the specific index structure in the triple-$K$ integral above. Conversely, triple-$K$ integrals conveniently resolve the complicated structure of linear dependencies in \cite{Davydychev:1992xr} leading to a more natural representation of the 3-point function. The relation between triple-$K$ integrals and 1-loop integrals in momentum space is summarised in appendix \ref{sec:momentum}.

Let us also comment on the validity of equation \eqref{e:red_scheme3}. The left-hand side diverges for $\alpha - \beta_t = -2n$, where $n$ is a non-negative integer. Nevertheless, it can still be used for regulated integrals with $\reg{\alpha} - \reg{\beta}_t = -2n + (u - v_t) \epsilon$ so long as $\epsilon \neq 0$.  If $n = 0$, however, \eqref{e:red_scheme3} evaluates only the divergent part of the left-hand side, assuming the integrals on the right-hand side are only known up to the finite part of order $\epsilon^0$. For example, \eqref{e:red_scheme3} relates the divergent part of $I_{\reg{2}\{\reg{1}\reg{1}\reg{1}\}}$ to the finite integral $I_{3\{011\}}$. It is therefore impossible to use \eqref{e:red_scheme3} to retrieve the finite part of $I_{\reg{2}\{\reg{1}\reg{1}\reg{1}\}}$ from a knowledge of $I_{3\{011\}}$.

\subsection{Reduction} \label{sec:scheme}

\begin{figure}[ht]
\centering
\includegraphics[width=0.80\textwidth]{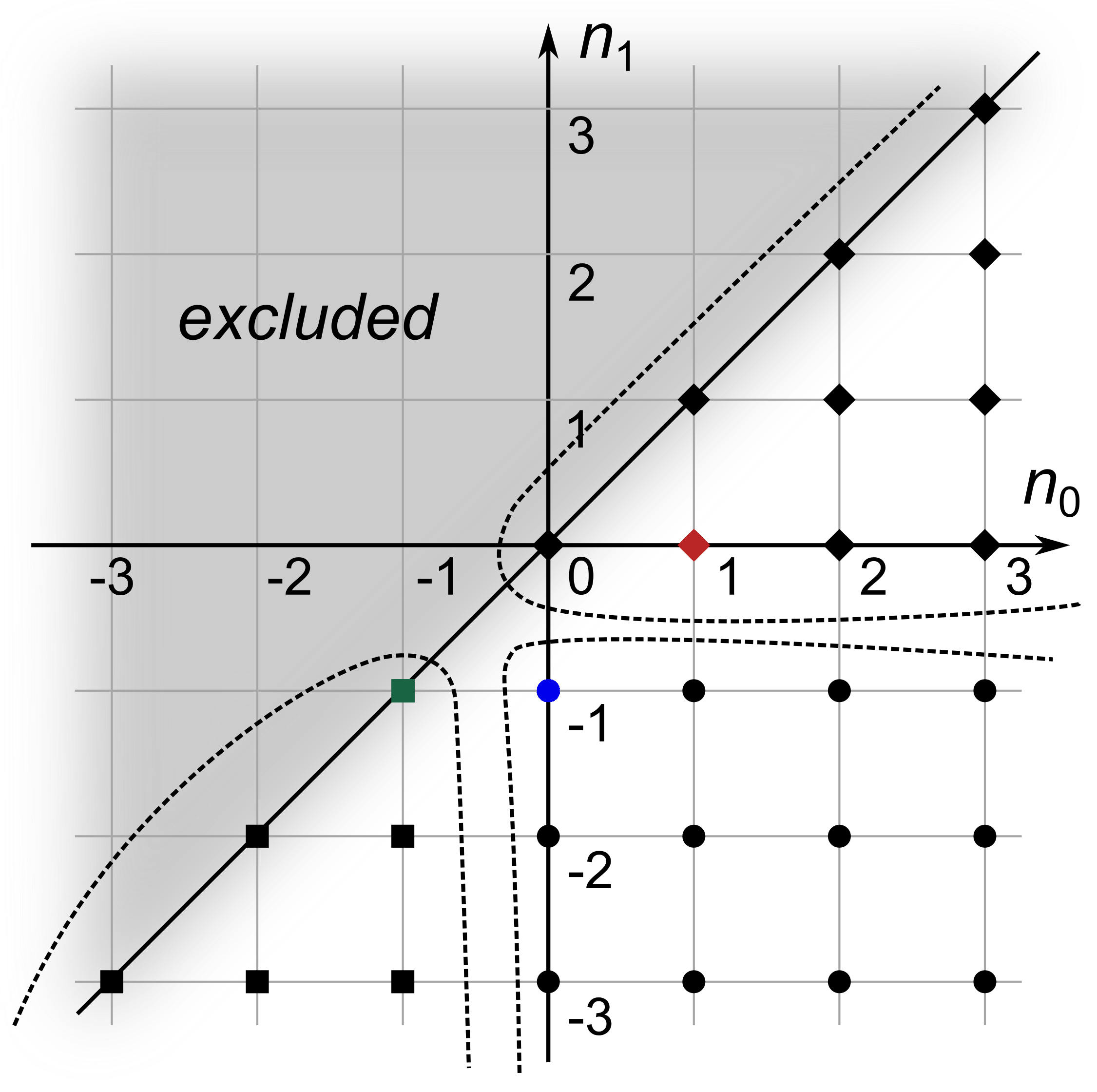}
\caption{A summary of all triple-$K$ integrals that can be obtained from the master integral $I_{0\{111\}}$ using the reduction scheme. Depending on the values of $n_0$ and $n_1$ the integrals are either finite (represented by squares), linearly divergent (circles), or quadratically divergent (diamonds). Each point represents an infinite series of integrals having the same values of $n_0$ and $n_1$. In particular the master integral $I_{0\{111\}}$ belongs to the series denoted by the red diamond at $(n_0,n_1)=(1,0)$. The linearly divergent integral $I_{2\{111\}}$ belongs to the series denoted by the blue circle at $(0,-1)$, while finite $I_{1\{000\}}$ to the series denoted by the green square at $(-1,-1)$. The three types of integrals are gathered into three separate regions denoted by dashed borders. Since $n_0 \geq n_1$, no integral can appear in the excluded region.}
\label{fig:g2}
\end{figure}

In this section we now present the complete reduction scheme. We will analyse cases in  order of increasing complexity. The integrals that can be evaluated through the scheme are presented graphically in figure \ref{fig:g2}.

\subsubsection{\texorpdfstring{Finite integrals ($n_0 < 0$)}{Finite integrals}}

Let us start with finite integrals satisfying $n_0 < 0$. We can simply use \eqref{e:red_scheme2} and \eqref{e:red_scheme23} in combination to write the reduction formula,
\begin{align}
&I_{\alpha \{ \beta_1 \beta_2 \beta_3\}}  = (-1)^{\beta_t} \K_{j, \beta_j}^{|n_0|-1} \left[ p_1^{2 \beta_1} p_2^{2 \beta_2} p_3^{2 \beta_3} \left( \frac{1}{p_1} \frac{\partial}{\partial p_1} \right)^{\beta_1} \left( \frac{1}{p_2} \frac{\partial}{\partial p_2} \right)^{\beta_2} \left( \frac{1}{p_3} \frac{\partial}{\partial p_3} \right)^{\beta_3} I_{1\{000\}} \right]
\end{align}
where $j$ can take any of the values $j=1,2,3$. It is crucial here that $n_0$ is strictly less than zero. This formula expresses any finite integral under consideration in terms of the finite $I_{1\{000\}}$ integral.

\subsubsection{\texorpdfstring{Linearly divergent integrals ($n_0 \geq 0$ and $n_1 < 0$)}{Linearly divergent integrals}}

Every linearly divergent integral under consideration satisfies $n_0 \geq 0$ and $n_1 < 0$, and can be reduced to the $I_{2\{111\}}$ integral. Notice that these restrictions imply $\beta_3 > 0$.  If $\beta_3 = 0$, we obtain $n_0 = n_1 \geq 0$, a contradiction.

The reduction procedure consists of two steps. First we reduce the triple-$K$ integral $I_{\tilde{\alpha} \{ \tilde{\beta}_1 \tilde{\beta}_2 \tilde{\beta}_3 \}}$ to an integral of the form $I_{\beta_3 + 1 \{\beta_3\beta_3 \beta_3 \}}$
with all beta parameters equal by means \\[0.5ex]
of \eqref{e:red_scheme2a} and \eqref{e:red_scheme23},
\begin{align} \label{e:fullreductionmmm}
I_{\tilde{\alpha} \{ \tilde{\beta}_1 \tilde{\beta}_2 \tilde{\beta}_3 \}} & = (-1)^{k_1 + k_2} \K_{j, \tilde{\beta}_j}^{|n_1|-1} \left[ p_1^{2 \tilde{\beta}_1} p_2^{2 \tilde{\beta}_2} \left( \frac{1}{p_1} \frac{\partial}{\partial p_1} \right)^{k_1} \left( \frac{1}{p_2} \frac{\partial}{\partial p_2} \right)^{k_2} \right.  \nn\\
& \qquad\qquad \left. \times \left( p_1^{- 2 \beta_3 - 2 v_1 \epsilon} p_2^{- 2 \beta_3 - 2 v_2 \epsilon} I_{\beta_3 + 1 + u \epsilon \{ \beta_3 + v_1 \epsilon, \beta_3 + v_2 \epsilon, \beta_3 + v_3 \epsilon \}} \right) \right],
\end{align}
where $j$ takes values $j=1,2,3$ and we defined
\begin{equation} \label{e:k1k2}
k_1 = \beta_1 - \beta_3, \qquad\qquad k_2 = \beta_2 - \beta_3.
\end{equation}

If $\beta_3 = 1$ then the integral on the right-hand side is the familiar $I_{2\{111\}}$ integral regularised in a generic scheme. If $\beta_3 > 1$, then the integral on the right-hand side has the form $I_{n+1\{nnn\}}$ with $n \geq 2$, and the equation \eqref{e:redbeta} can be used recursively. One has
\begin{equation}
I_{n + 1 + u \epsilon \{ n + v_1 \epsilon, n + v_2 \epsilon, n + v_3 \epsilon \}} = \frac{\B_{n + v_1 \epsilon, n + v_2 \epsilon, n + v_3 \epsilon} I_{n + u \epsilon \{ n - 1 + v_1 \epsilon, n - 1 + v_2 \epsilon, n - 1 + v_3 \epsilon \}}}{-2 n + 2 + \epsilon(u - v_t)},
\end{equation}
where the operator $\B$ is defined in \eqref{e:Bbeta}. The recursion is well defined since $n \geq 2$.

This completes the reduction of the integral to $I_{2 + u \ep\{1 + v_1 \ep, 1 + v_2 \ep, 1 + v_3 \ep\}}$. Notice that at all stages of the calculation one needs to keep track of subleading terms in the regulator $\epsilon$. Indeed, since the integrals are divergent, subleading terms in $\epsilon$ may combine with divergent pieces producing additional contributions to the finite part.

\subsubsection{\texorpdfstring{Integrals with $n_1 = 0$}{Integrals with n1=0}}

This interesting class of integrals 
arises for 3-point functions of marginal scalar operators in even-dimensional CFTs. The triple-$K$ integral exhibits a double pole and can be reduced to the master integral $I_{0\{111\}}$. The method is almost identical to that presented in the previous subsection.

First, however, consider the special case  $n_0 = 0$, which implies $\beta_3 = 0$. Here, we have the relation 
\begin{align}
I_{\tilde{\alpha}\{\tilde{\beta}_1 \tilde{\beta}_2 0\}} & = (-1)^{\beta_1 + \beta_2 + 1} p_1^{2 \tilde{\beta}_1} p_2^{2 \tilde{\beta}_2} \left( \frac{1}{p_1} \frac{\partial}{\partial p_1} \right)^{\beta_1 - 1} \left( \frac{1}{p_2} \frac{\partial}{\partial p_2} \right)^{\beta_2 - 1} \frac{1}{p_3} \frac{\partial}{\partial p_3} \nn\\
& \qquad\qquad \times \left[ p_1^{-2 - 2 v_1 \epsilon} p_2^{-2 - 2 v_2 \epsilon} I_{0 + u \ep \{1 + v_1 \ep, 1 + v_2 \ep, 1 + v_3 \ep\}} \right],
\end{align}
where it is important that $\beta_1 \geq \beta_2 \geq 1$. Indeed, if in addition to $\beta_3 = 0$ we  also had $\beta_2 = 0$, \eqref{e:n2def} would then imply $n_2 = 0$ contradicting our assumption  $n_2 < 0$.

Consider now the more general case where $n_1 = 0$ and $n_0 > 0$, implying $\beta_3 > 0$. In this case we can use a modification of \eqref{e:fullreductionmmm} to write
\begin{align} \label{e:fullreduction}
I_{\tilde{\alpha} \{ \tilde{\beta}_1 \tilde{\beta}_2 \tilde{\beta}_3 \}} & = (-1)^{k_1 + k_2} p_1^{2 \tilde{\beta}_1} p_2^{2 \tilde{\beta}_2} \left( \frac{1}{p_1} \frac{\partial}{\partial p_1} \right)^{k_1} \left( \frac{1}{p_2} \frac{\partial}{\partial p_2} \right)^{k_2}  \nn\\ 
& \qquad\qquad \times \left[ p_1^{- 2 \beta_3 - 2 v_1 \ep} p_2^{- 2 \beta_3 - 2 v_2 \ep} I_{\beta_3 - 1 + u \epsilon \{ \beta_3 + v_1 \ep, \beta_3 + v_2 \ep, \beta_3 + v_3 \ep \}} \right],
\end{align}
where $k_1$ and $k_2$ are defined in \eqref{e:k1k2}. For $\beta_3 = 1$ the integral on the right-hand side is the master integral $I_{0\{111\}}$ regularised in a generic scheme. If $\beta_3 > 1$, the integral on the right-hand side has the form $I_{n-1\{nnn\}}$ with $n \geq 1$, and as previously \eqref{e:redbeta} can be used recursively.  We find
\begin{equation}
I_{n-1 + u \epsilon \{ n + v_1 \epsilon, n + v_2 \epsilon, n + v_3 \epsilon \}} = \frac{\B_{n + v_1 \epsilon, n + v_2 \epsilon, n + v_3 \epsilon} I_{n - 2 + u \epsilon \{ n - 1 + v_1 \epsilon, n - 1 + v_2 \epsilon, n - 1 + v_3 \epsilon \}}}{-2 n + \epsilon(u - v_t)},
\end{equation}
where the operator $\B$ was defined in \eqref{e:Bbeta}. The recursion is well defined since $n \geq 1$.

\subsubsection{\texorpdfstring{Integrals with $n_1 > 0$}{Integrals with n1>0}}

The reduction procedure for integrals with $n_1 > 0$ is more involved.  
By using \eqref{e:red_scheme3} recursively, however, one can eventually express all such integrals in terms of the master integral $I_{0\{111\}}$.

We will consider two cases.  The first consists of all integrals satisfying $n_0 > n_1 > 0$, \textit{i.e.}, integrals lying in the interior of the upper-right wedge in figure \ref{fig:g2}. The second case consists of integrals satisfying $n_0 = n_1 > 0$, \textit{i.e.}, integrals lying on the diagonal line in figure \ref{fig:g2}. 

We will show that in the first case the application of \eqref{e:red_scheme3} leads to a sum of integrals with the values of their corresponding constants $n_0$ decreased by one. Similarly, in the second case, we will show that the original integral can be re-expressed as a combination of triple-$K$ integrals with values of $n_1$ that have been decreased by one. In both cases the recursion is such that values of $n_2$ remain negative throughout.

In each step of the reduction procedure the values of $n_0$ and $n_1$ are thus decreased until we reach the point where $n_1 = 0$, allowing the results of the previous subsection to be utilised.

\bigskip \noindent \textbf{Case $\bs{n_0 > n_1}$.} If $n_0 > n_1$ we use equation \eqref{e:red_scheme3}, which can be written as
\begin{align}
I_{\tilde{\alpha} \{ \tilde{\beta}_1 \tilde{\beta}_2 \tilde{\beta}_3 \}} & = \frac{1}{- 2 n_0 + (u - v_t) \epsilon} \left[ p_1^2 I_{\tilde{\alpha} + 1 \{\tilde{\beta}_1 - 1, \tilde{\beta}_2, \tilde{\beta}_3 \}} + p_2^2 I_{\tilde{\alpha} + 1 \{\tilde{\beta}_1, \tilde{\beta}_2 - 1, \tilde{\beta}_3 \}}  
\right.\nn\\& \qquad\qquad\qquad\qquad\qquad \left. 
+ \: p_3^2 I_{\tilde{\alpha} + 1 \{\tilde{\beta}_1, \tilde{\beta}_2, \tilde{\beta}_3 - 1\}} \right]. \label{e:red_scheme3_copy}
\end{align}
Due to the ordering \eqref{e:order_beta} we have $n_0 \geq n_1 > 0$, and hence the use of \eqref{e:red_scheme3_copy} is always justified.

Assume that a given integral on the left-hand side of \eqref{e:red_scheme3_copy} possesses the associated  constants $n_0, n_1, n_2$. We want to calculate the values of the corresponding constants -- say $n_0', n_1', n_2'$ -- for the integrals on the right-hand side, and to confirm that they decrease.

Indeed, $n_0 > n_1$ is equivalent to $\beta_3 > 0$, and for the integrals on the right-hand side we then have
\begin{center}
\begin{tabular}{|c|c|c|c|}
\hline
Integral & $n'_0$ & $n'_1$ & $n'_2$ \\ \hline
$I_{\tilde{\alpha} + 1 \{\tilde{\beta}_1 - 1, \tilde{\beta}_2, \tilde{\beta}_3 \}}$ & $n_0 - 1$ & $n_1 - 1$ & $n_2 - 1$ \\ \hline
$I_{\tilde{\alpha} + 1 \{\tilde{\beta}_1, \tilde{\beta}_2 - 1, \tilde{\beta}_3 \}}$ & $n_0 - 1$ & $n_1 - 1$ & $n_2$ \\ \hline
$I_{\tilde{\alpha} + 1 \{\tilde{\beta}_1, \tilde{\beta}_2, \tilde{\beta}_3 - 1\}}$ & $n_0 - 1$ & $n_1$ & $n_2$ \\ \hline
\end{tabular}
\end{center}
As we can see $n_2' \leq n_2 < 0$ and in all cases the value of $n_0$ decreases by one.  The repeated use of the equation \eqref{e:red_scheme3_copy} will thus lead to a combination of integrals satisfying either $n_1 = 0$ or $n_0 = n_1$ with $n_2 < 0$.

\bigskip \noindent \textbf{Case $\bs{n_0 = n_1}$.} The remaining case is the analysis of $n_0 = n_1 > 0$. This, however, implies that $\beta_3 = 0$ and the last integral in \eqref{e:red_scheme3_copy} therefore has a negative value of its $\beta_3$ coefficient. In this case we can flip the sign using \eqref{e:2id} leading to a modification of \eqref{e:red_scheme3_copy} for $\beta_3 = 0$, 
\begin{align}
I_{\tilde{\alpha} \{ \tilde{\beta}_1 \tilde{\beta}_2, 0 + v_3 \epsilon \}} & = \frac{1}{- 2 n_0 + (u - v_t) \epsilon} \left[ p_1^2 I_{\tilde{\alpha} + 1 \{\tilde{\beta}_1 - 1, \tilde{\beta}_2, v_3 \epsilon \}} + p_2^2 I_{\tilde{\alpha} + 1 \{\tilde{\beta}_1, \tilde{\beta}_2 - 1, v_3 \epsilon \}} \right.\nn\\ 
& \qquad\qquad \qquad\qquad\qquad\left. + \: p_3^{2 v_3 \epsilon} I_{\tilde{\alpha} + 1 \{\tilde{\beta}_1, \tilde{\beta}_2, 1-v_3 \epsilon\}} \right]. \label{e:red_scheme3_copy_mod}
\end{align}
The last integral on the right-hand side satisfies $n_0' = n_0$, $n_1' = n_1 - 1$ and $n_2' = n_2 - 1$.  All integrals on the right-hand side of \eqref{e:red_scheme3_copy_mod} thus have their values of $n_1$ decreased by one and satisfy $n_0' \geq n_1' \geq 0$. This concludes the reduction procedure.


\subsection{Tensorial operators in $d = 4$} \label{sec:tensors}

3-point functions involving the stress tensor and conserved currents in four-dimensional CFT are of special interest. A complete set of unrenormalised momentum-space expressions for such correlation functions was presented in \cite{Bzowski:2013sza}. All required integrals fit into the diagram presented in table \ref{e:table}.

The integrals in table \ref{e:table} can all be obtained as indicated from a single master integral, $I_{0\{111\}}$, whose value is given in \eqref{e:masterap}. Where a `cell' of the diagram contains two entries the corresponding arrows carry two operators: of these two operators, the upper one leads to/is applied to the upper entry and lower operator to the lower entry.

All integrals are implicitly assumed to be regulated in a single but arbitrary scheme with fixed values of $u$ and $v_j$.  Calculating the parameters $n_0$ and $n_1$, equations \eqref{e:n0def} and \eqref{e:n1def} then identify the degree of divergence of all integrals in the table. \red{Red} entries then indicate integrals exhibiting a double pole in the regulator; \navy{blue} entries those that are linearly divergent; and \green{green}, finite integrals. 

The operators $L_j$ and $M_j$ are the differential operators featuring in \eqref{e:1id} and \eqref{e:red_scheme2}, defined as
\begin{equation}
L_j = - \frac{1}{p_j} \frac{\partial}{\partial p_j}, \qquad\qquad M_j = 2 \reg{\beta}_j - p_j \frac{\partial}{\partial p_j}.
\end{equation}
The corresponding regulated value $\reg{\beta}_j$ is equal to that of the integral on which $M_j$ is acting. The dotted line in table \ref{e:table} indicates the use of \eqref{e:red_scheme3}.

\begin{center}
\begin{table}[t]
\hspace{0.0cm}
\xymatrix@R+=60pt@C+=60pt{
*+[F]{\red{I_{0\{111\}}}} \ar[r]^{M_1} & \red{I_{1\{211\}}} \ar[r]^{L_1} \ar[d]^{M_2} & \navy{I_{2\{111\}}} \ar[d]^{M_1} & \\
\red{I_{1\{222\}}} \ar[r]^{L_3} \ar[d]^{M_1} & \red{I_{2\{221\}}} \ar@{.>}@/^1pc/[l] \ar[r]^{L_2} \ar[d]^{\substack{M_3 \\ M_1}} & \navy{I_{3\{211\}}} \ar[r]^{L_1} \ar[d]^{\substack{M_2 \\ M_1}} & \green{I_{4\{111\}}} \ar[d]^{M_1} \\
\red{I_{2\{322\}}} \ar[r]^{L_1}_{L_3} \ar[d]^{M_2} & {\begin{array}{c} \navy{I_{3\{222\}}} \\ \red{I_{3\{321\}}} \end{array}} \ar[r]^{L_3}_{L_2} \ar[d]^{\substack{M_1 \\ M_3}} & {\begin{array}{c} \navy{I_{4\{221\}}} \\ \navy{I_{4\{311\}}} \end{array}} \ar[r]^{L_2}_{L_1} \ar[d]^{\substack{M_3 \\ M_2}} & \green{I_{5\{211\}}} \ar[d]^{M_2} \\
\red{I_{3\{332\}}} \ar[r]^{L_2} \ar[d]^{\substack{M_3 \\ M_1} } & \navy{I_{4\{322\}}} \ar[r]^{L_1}_{L_3} \ar[d]^{\substack{M_2 \\ M_1}} & {\begin{array}{c} \navy{I_{5\{222\}}} \\ \navy{I_{5\{321\}}} \end{array}} \ar[r]^{L_3}_{L_1} \ar[d]^{\substack{M_1 \\ M_3}} & \green{I_{6\{221\}}} \ar[d]^{M_3} \\
{\begin{array}{c} \navy{I_{4\{333\}}} \\ \red{I_{4\{432\}}} \end{array}} \ar[r]^{L_3}_{L_2} & {\begin{array}{c} \navy{I_{5\{332\}}} \\ \navy{I_{5\{422\}}} \end{array}} \ar[r]^{L_2}_{L_1} & \navy{I_{6\{322\}}} \ar[r]^{L_1} & \green{I_{7\{222\}}}
}
\centering
\caption{Reduction scheme for the integrals required to calculate all 3-point functions of conserved currents and the stress tensor in $d=4$. The operators $L_j$ and $M_j$ are defined by $L_j = -p_j^{-1} \partial/\partial p_j$ and $M_j = 2 \reg{\beta}_j - p_j \partial/\partial p_j$.  Further details may be found in section \ref{sec:tensors}.}
\label{e:table}
\end{table}
\end{center}

\vspace{-10mm}

\section{Master integral}  \label{sec:MI}

In this section we derive the expression for the master integral $I_{0\{111\}}$, or rather its regularised version \eqref{e:iireg},
\begin{equation}
I_{0 + u \ep \{ 1 + v_1 \ep, 1 + v_2 \ep, 1 + v_3 \ep\}},
\end{equation}
where $u$ and $v_j$, $j=1,2,3$ are fixed but otherwise arbitrary. As we showed in the previous section, every triple-$K$ integral satisfying conditions (a) - (c) from section \ref{sec:review} can be expressed in terms of the master integral $I_{0\{111\}}$. For convenience we will also present expressions for the two auxiliary master integrals featured in figure \ref{fig:g2}, $I_{1\{000\}}$ and $I_{2\{111\}}$.

Our evaluation of the master integral is based upon the relation between triple-$K$ integrals and hypergeometric functions. 
In the context of 1-loop 3-point integrals in momentum space similar relations have been analysed in a number of papers, for example \cite{CabralRosetti:1998sp,Davydychev:1999mq,Anastasiou:1999ui,Boos:1987bg,Davydychev:2005nf,CabralRosetti:2002qv,Kniehl:2011ym,Coriano:2013jba}.  The conformal case, however, corresponds to massless integrals and hence the resulting expressions can usually be simplified to more elementary functions. The earliest examples containing dilogarithms can be traced back to \cite{Passarino:1978jh,tHooft:1978xw}.

\subsection{The result}

The master integral takes the following form
\begin{equation} \label{e:masterap}
I_{0+u \epsilon, \{1 + v_1 \epsilon, 1 + v_2 \epsilon, 1 + v_3 \epsilon\}} = \frac{I^{(-2)}}{\epsilon^2} + \frac{I^{(-1)}}{\epsilon} + I^{(\text{scheme})} + I^{(\text{non-local})} + I^{(\text{scale-violating})} + O(\epsilon),
\end{equation}
where
\begin{align}
& I^{(-2)} = \frac{1}{2 (v_t - u)} \sum_{j=1}^3 \frac{p_j^2}{u - v_t + 2 v_j}, \\
& I^{(-1)} = \frac{1}{4} \sum_{j=1}^3 \frac{p_j^2 \log p_j^2}{u - v_t + 2 v_j} + \frac{u(1 - 2 \gamma_E + 2 \log 2) - v_t}{4(v_t - u)} \sum_{j=1}^3 \frac{p_j^2}{u - v_t + 2 v_j}, \\
& I^{(\text{scheme})} = \frac{1}{8} \sum_{j=1}^3 \frac{v_j}{u - v_t + 2 v_j} p_j^2 \log^2 p_j^2 + \frac{1}{8} \left[ u(1 - 2 \gamma_E + 2 \log 2) - v_t \right] \sum_{j=1}^3 \frac{p_j^2 \log p_j^2}{u - v_t + 2 v_j} \nn\\
& \qquad\qquad + \: \frac{\left[ v_t - u(1 - \gamma_E + \log 2) \right]^2 + u^2 (\gamma_E - \log 2)^2 + \tfrac{1}{3} \pi^2 v_{tt}}{8(v_t - u)} \sum_{j=1}^3 \frac{p_j^2}{u - v_t + 2 v_j}, \label{e:Ischeme} \\
& I^{(\text{non-local})} = - \frac{1}{8} \sqrt{-J^2} \left[ \frac{\pi^2}{6} - 2 \log \frac{p_1}{p_3} \log \frac{p_2}{p_3} + \log X \log Y - \Li_2 X  - \Li_2 Y \right], \label{e:intnonloc} \\
& I^{(\text{scale-violating})} = \frac{1}{16} \left[ (p_3^2 - p_1^2 - p_2^2) \log p_1^2 \log p_2^2 + (p_2^2 - p_1^2 - p_3^2) \log p_1^2 \log p_3^2 \right.\nn\\
& \qquad\qquad\qquad\qquad\qquad \left. + \: (p_1^2 - p_2^2 - p_3^2) \log p_2^2 \log p_3^2 \right].
\end{align}
The meaning of various parts is as follows:
\begin{itemize}
\item $I^{(-2)}$ and $I^{(-1)}$ are the coefficients of the divergent parts of the integrals. These can be extracted straightforwardly using the method of section \ref{sec:extractdiv}. 
\item $I^{(\text{scheme})}$ denotes the finite part of the integral depending on the regularisation parameters $u$ and $v_j$. (Note that the coefficients $I^{(-2)}$ and $I^{(-1)}$ depend on $u$ and $v_j$ as well.) The remaining pieces $I^{(\text{non-local})}$ and $I^{(\text{scale-violating})}$ do not depend on either $u$ or $v_j$.
\item $I^{(\text{non-local})}$ contains the essential non-local part of the integral, as well as the only special function, the dilogarithm $\Li_2$. This part is scale-invariant, \textit{i.e.}, all logarithms depend on the ratios $p_1/p_3$ and $p_2/p_3$ only.  
\item $I^{(\text{scale-violating})}$ contains the non-local yet scale-violating part of the integral. 
After renormalisation, these terms are related to beta functions and conformal anomalies.
\end{itemize}
In writing the master integral, we used the definitions
\begin{align}
J^2 & = (p_1 + p_2 - p_3) (p_1 - p_2 + p_3) (-p_1 + p_2 + p_3) (p_1 + p_2 + p_3), \label{e:J2} \\
X & = \frac{- p_1^2 + p_2^2 + p_3^2 - \sqrt{-J^2}}{2 p_3^2}, \qquad Y = \frac{- p_2^2 + p_1^2 + p_3^2 - \sqrt{-J^2}}{2 p_3^2}, \label{e:XY} \\
v_t & = v_1 + v_2 + v_3, \qquad\qquad v_{tt} = v_1^2 + v_2^2 + v_3^2.
\end{align}
The decomposition we have made is by no means unique. It simply organises the final result neatly. We discuss further properties of the master integral in the following section.

\subsection{Tools and identities}

The physical implications of the double logarithms of momenta in the scale-violating part of the master integral were discussed in \cite{Scalars}. In this section we will concentrate on the `non-local' part $I^{\text{(non-local)}}$ containing the dilogarithms.

The quantity $J^2$ defined in \eqref{e:J2} has a geometric interpretation. 
Physically, the $p_j = | \bs{p}_j |$ are the magnitudes of three $d$-dimensional vectors $\bs{p}_1, \bs{p}_2, \bs{p}_3$
satisfying $\bs{p}_1 + \bs{p}_2 + \bs{p}_3 = 0$ due to momentum conservation. 
Scalar products between these momentum vectors can be expressed in terms of the $p_j$ according to
\begin{equation}
\bs{p}_1 \cdot \bs{p}_2 = \frac{1}{2} ( p_3^2 - p_1^2 - p_2^2)
\end{equation}
and similarly for other products. Equation \eqref{e:J2} can then be rewritten as
\begin{align}
J^2 & = (p_1 + p_2 - p_3) (p_1 - p_2 + p_3) (-p_1 + p_2 + p_3) (p_1 + p_2 + p_3), \nn\\
& = -p_1^4 - p_2^4 - p_3^4 + 2 p_1^2 p_2^2 + 2 p_1^2 p_3^2 + 2 p_2^2 p_3^2 \nn\\
& = 4 \left[ p_1^2 p_2^2 - (\bs{p}_1 \cdot \bs{p}_2)^2 \right] = 4 \cdot \text{Gram}(\bs{p}_1, \bs{p}_2),
\end{align}
where $\text{Gram}$ is the Gram determinant.  The area of a triangle with side lengths $p_1$, $p_2$ and $p_3$ is therefore $(1/4)\sqrt{J^2}$.
For physical momentum configurations obeying the triangle inequalities one has $J^2 \geq 0$, with $J^2 = 0$ holding if and only if the momenta $\bs{p}_j$ are collinear.

The definitions of $X$ and $Y$ in \eqref{e:XY} agree with those in \cite{Davydychev:1992xr}.
In the literature, \textit{e.g.}, \cite{CabralRosetti:1998sp,Duhr:2014woa}, an alternative choice of  variables in place of $X$ and $Y$ is commonly encountered; usually these are denoted by $z$ and its complex conjugate $\bar{z}$, defined as any pair of solutions to the quadratic equations
\begin{equation}
z \bar{z} = \frac{p_2^2}{p_3^2}, \qquad\qquad (1 - z)(1 - \bar{z}) = \frac{p_1^2}{p_3^2}.
\end{equation}
The relation between $z, \bar{z}$ and $X, Y$ is then simply $z = X$ and $\bar{z} = 1 - Y$.
In such a representation the `non-local' part of the master integral thus reads 
\begin{equation}
I^{(\text{non-local})} = \frac{1}{8} (\bar{z} - z) p_3^2 \left[ \frac{1}{2} \log (z \bar{z}) \left( \log(1 - z) - \log(1 - \bar{z}) \right) + \Li_2 z - \Li_2 \bar{z} \right].
\end{equation}

Unfortunately it is not obvious in either representation that $I^{\text{(non-local)}}$ is real and symmetric under any permutation of momenta $p_j$, $j=1,2,3$. Nevertheless, the function is indeed completely symmetric by virtue of certain identities between dilogarithms, see for example \cite{Erdelyi,Anastasiou:1999ui}.

Every integral which can be obtained from the master integral by means of the reduction scheme discussed in section \ref{sec:eval} contains a piece proportional to $I^{\text{(non-local)}}$. It is then rather convenient to define the quantity 
\begin{align} \label{e:L}
\mathcal{L} & = \frac{\pi^2}{6} - 2 \log \frac{p_1}{p_3} \log \frac{p_2}{p_3} + \log X \log Y - \Li_2 X - \Li_2 Y \nn\\
& = - \frac{1}{2} \log (z \bar{z}) \left( \log(1 - z) - \log(1 - \bar{z}) \right) - \Li_2 z + \Li_2 \bar{z}.
\end{align}
$\mathcal{L}$ is purely imaginary and completely symmetric under any permutation of $p_j$, $j=1,2,3$, and moreover derivatives of $\mathcal{L}$ turn out to be extremely simple, 
\begin{align}
\frac{\partial \mathcal{L}}{\partial p_1} & = \frac{2}{p_1 \sqrt{-J^2}} \left[ p_1^2 \log p_1^2 + \frac{1}{2} (p_3^2 - p_1^2 - p_2^2) \log p_2^2 + \frac{1}{2} (p_2^2 - p_1^2 - p_3^2) \log p_3^2 \right] \nn\\
& = \frac{2}{p_1 \sqrt{-J^2}} \left[ p_1^2 \log p_1^2 + \bs{p}_1 \cdot \bs{p}_2 \log p_2^2 + \bs{p}_1 \cdot \bs{p}_3 \log p_3^2 \right], \label{e:dL}
\end{align}
where in the last line we used the physical interpretation of the parameters $p_j$ as magnitudes of vectors $\bs{p}_j$ satisfying $\bs{p}_1 + \bs{p}_2 + \bs{p}_3 = 0$. To evaluate  derivatives of $\mathcal{L}$ with respect to other $p_j$ one simply needs to permute the momenta accordingly.

\subsection{Auxiliary integrals}

In the course of the reduction scheme, we expressed all integrals satisfying conditions (a) - (c) from section \ref{sec:review} in terms of three integrals: $I_{0\{111\}}$, $I_{2\{111\}}$ and $I_{1\{000\}}$. The latter two auxiliary integrals are related to the master integral $I_{0\{111\}}$ by the identities \eqref{e:redbeta} and \eqref{e:red_scheme23} respectively, 
\begin{align}
& I_{1\{000\}} = \frac{1}{2 p_1 p_2 p_3} \left[ p_1 \frac{\partial^2}{\partial p_2 \partial p_3} + p_2 \frac{\partial^2}{\partial p_1 \partial p_3} + p_3 \frac{\partial^2}{\partial p_1 \partial p_2} \right] I_{0+u \epsilon\{1 + v \epsilon, 1 + v \epsilon, 1 + v \epsilon\}} + O(\epsilon), \label{e:I1000from_master}\\
& I_{2+u \epsilon\{1 + v_1 \epsilon, 1 + v_2 \epsilon, 1 + v_3 \epsilon\}} = \K_{j, 1+v_j \epsilon} I_{0+u \epsilon\{1 + v_1 \epsilon, 1 + v_2 \epsilon, 1 + v_3 \epsilon\}}, \label{e:I2111from_master}
\end{align}
where the operator $\K_{j, 1 + v_j \epsilon} = \K_j$ is the conformal Ward identity operator given in \eqref{e:KOp} and $j$ takes an arbitrary value $j=1,2,3$.  The first integral is finite and hence does not require a regulator. The second integral exhibits a single pole. Using the definition \eqref{e:L} for $\mathcal{L}$, their exact expressions can be derived and read
\begin{align}
I_{1 \{000\}} & = \frac{\mathcal{L}}{2 \sqrt{-J^2}}, \label{e:I1000} \\
I_{2 + u \epsilon \{1 + v_1 \epsilon, 1 + v_2 \epsilon, 1 + v_3 \epsilon \}} & = \frac{1}{(u - v_t) \epsilon} + \frac{2 p_1^2 p_2^2 p_3^2}{(-J^2)^{3/2}} \mathcal{L} + \frac{u}{u - v_t} ( \log 2 - \gamma_E ) \nn\\
& \qquad - \: \frac{1}{2 J^2} \left[ p_1^2 (p_2^2 + p_3^2 - p_1^2) \log p_1^2 + p_2^2 (p_1^2 + p_3^2 - p_2^2) \log p_2^2 \right.\nn\\
& \qquad\qquad\qquad \left. + \: p_3^2 (p_1^2 + p_2^2 - p_3^2) \log p_3^2 \right] + O(\epsilon), \label{e:I2111}
\end{align}
The first integral $I_{1\{000\}}$ is known in the literature (see \textit{e.g.}, \cite{Boos:1987bg,Davydychev:1992xr,tHooft:1978xw}) and represents, for example, the 3-point function of $\phi^2$ in the 4-dimensional theory of free massless scalars.

\subsection{Evaluation} \label{sec:master_eval}

To evaluate the master integral we used a sequence of mathematical identities.
This sequence, to be explained in the following subsections, is:
\begin{enumerate}
\item We first evaluate integrals of the form $I_{\nu+1\{\nu\nu\nu\}}$ for any $\nu \in \R$.  Integrals of this form can be expressed in terms of hypergeometric functions (Legendre functions).
\item Substituting $\nu=-1+\epsilon$ and expanding in $\epsilon$, we evaluate the triple-$K$ integral $I_{0 + \epsilon \{ -1+\epsilon, -1+\epsilon, -1+\epsilon \}}$. The result contains a single special function, the dilogarithm.
\item Using \eqref{e:2id}, we obtain 
\begin{equation} \label{e:how_to_get_master}
I_{0 + \epsilon \{ 1-\epsilon, 1-\epsilon, 1-\epsilon \}} = (p_1 p_2 p_3)^{2 - 2 \epsilon} I_{0 + \epsilon \{ -1+\epsilon, -1+\epsilon, -1+\epsilon \}}
\end{equation}
which is the master integral in a regularisation scheme with $-u = v_1 = v_2 = v_3$.
\item Finally, we change to an arbitrary regularisation scheme specified by general  $u$ and $v_j$ parameters according to the method described in section \ref{sec:schemechangedetails}.
\end{enumerate}

\subsubsection{\texorpdfstring{Evaluation of $I_{\nu+1\{\nu\nu\nu\}}$}{Evaluation of I_(v+1(vvv))}} \label{ch:integral}

To evaluate $I_{\nu+1\{\nu\nu\nu\}}$ we start with the representation of the triple-$K$ integral in terms of the generalised hypergeometric function  Appell $F_4$  \cite{Erdelyi,Appell},
\begin{align}
I_{\alpha \{\beta_1 \beta_2 \beta_3\}} = \frac{2^{\alpha - 4}}{p_3^\alpha} \left[ A(\lambda, \mu) + A(\lambda, -\mu) + A(-\lambda, \mu) + A(-\lambda, -\mu) \right], \label{e:KKK}
\end{align}
where
\begin{align}
A(\lambda, \mu) & = \left( \frac{p_1}{p_3} \right)^\lambda \left( \frac{p_2}{p_3} \right)^\mu \Gamma \left( \frac{\alpha + \lambda + \mu - \nu}{2} \right) \Gamma \left( \frac{\alpha + \lambda + \mu + \nu}{2} \right) \Gamma(-\lambda) \Gamma(-\mu)  \nn\\
& \qquad \times F_4 \left( \frac{\alpha + \lambda + \mu - \nu}{2}, \frac{\alpha + \lambda + \mu + \nu}{2}; \lambda + 1, \mu + 1; \frac{p_1^2}{p_3^2}, \frac{p_2^2}{p_3^2} \right).
\end{align}
This representation is not very useful for numerical evaluation, but provides a good starting point for formal manipulations. For triple-$K$ integrals of the form $I_{\nu+1\{\nu\nu\nu\}}$, the Appell function simplifies to regular hypergeometric functions. Using the reduction formulae \eqref{e:redform1} to \eqref{e:redform2}, we find
\begin{align}
I_{\nu+1 \{\nu\nu\nu\}} & = \frac{2^{\nu-2} \Gamma(\nu) \pi}{\sin(\pi \nu)} \left[ \frac{p_3^{2 + 2 \nu}}{p_1^2 p_2^2} X Y \: F_\nu \left( \frac{p_3^4}{p_1^2 p_2^2} X^2 Y^2 \right) - \frac{p_2^{2 \nu}}{p_1^2} Y \: F_\nu \left( \frac{p_3^2}{p_1^2} Y^2 \right) \right.\nn\\
& \qquad\qquad \left. - \: \frac{p_1^{2 \nu}}{p_2^2} X \: F_\nu \left( \frac{p_3^2}{p_2^2} X^2 \right) \right] + \frac{2^{3\nu - 2} \pi^{\frac{3}{2}} \Gamma \left( \nu + \frac{1}{2} \right)}{\sin^2(\pi \nu)} (p_1 p_2 p_3)^{2 \nu} ( \sqrt{-J^2} )^{-(2 \nu+1)}, \label{e:Jn}
\end{align}
where
\begin{equation} \label{e:Fnu}
F_\nu(x) = {}_2 F_1(1, \nu+1; 1-\nu; x)
\end{equation}
and the variables $X, Y$ are as defined in \eqref{e:XY} while $J^2$ is given in \eqref{e:J2}. Note that this particular combination of parameters in the hypergeometric function also appears in Legendre functions.

In terms of the parameters $n_0, n_1, n_2$ defined in \eqref{e:n0def} to \eqref{e:n2def}, any integral of the form $I_{\nu + 1 \{ \nu \nu \nu \}}$ with integer $\nu$ satisfies $n_0 = \nu - 1$ and $n_1 = - 1$ with $n_2 = - \nu - 1$.  Such integrals therefore lie on the horizontal line  $n_1 = -1$ in figure \ref{fig:g2} provided $n_2 < 0$. Note that the master integral is not in this class.
Nevertheless, by using the inversion trick \eqref{e:2id} we can write
\begin{equation}
I_{-\nu+1 \{\nu\nu\nu\}} = (p_1 p_2 p_3)^{2\nu} I_{-\nu+1 \{-\nu-\nu-\nu\}}.
\end{equation}
The integral on the right-hand side can be expressed through \eqref{e:Jn}, while the integral on the left-hand side satisfies $n_0 = 2 \nu - 1$, $n_1 = \nu - 1$ and $n_2 = -2 < 0$.  By this method we can then directly generate all integrals lying on the line $n_1 = (n_0 - 1) / 2$ in figure \ref{fig:g2} without use of the reduction scheme.

\subsubsection{Master integral}

For generic values of $\nu$ the expression \eqref{e:Jn} is finite. However, we are interested in $\nu = n + \epsilon$ close to an integer value of $n$ where the expression becomes singular. In such cases \eqref{e:Jn} can be series expanded around $\nu = n$ up to terms vanishing as $\ep\rightarrow 0$.
The expansion can be obtained by representing the hypergeometric function in terms of its usual power series,
\begin{equation} \label{e:Fser}
F_\nu(x) = \frac{\Gamma(1 - \nu)}{\Gamma(1 + \nu)} \sum_{k=0}^\infty \frac{\Gamma(1 + \nu + k)}{\Gamma(1 - \nu + k)} x^k.
\end{equation}
In the case of interest, $\nu = -1 + \epsilon$, the relevant expansion is
\begin{equation} \label{e:Fexp}
F_{-1+\epsilon}(x) = 1 + F^{(1)}_{-1}(x) \epsilon + F^{(2)}_{-1}(x) \epsilon^2 + O(\epsilon^3),
\end{equation}
where the expansion coefficients can be resummed as
\begin{align}
F^{(1)}_{-1}(x) & = 1 - \left( 1 - \frac{1}{x} \right) \log (1-x), \\
F^{(2)}_{-1}(x) & = 2 + \left( 1 - \frac{1}{x} \right) \left[ -\log(1-x) + \log^2(1-x) + \Li_2 x \right].
\end{align}

Combining everything we obtain an analytic expression for $I_{0+\epsilon \{-1+\epsilon, -1+\epsilon, -1+\epsilon\}}$ up to terms vanishing as $\epsilon \rightarrow 0$. 
Using \eqref{e:2id}, we then arrive at
\begin{equation} \label{e:i0111fromm}
I_{0+\epsilon \{1-\epsilon, 1-\epsilon, 1-\epsilon\}} = (p_1 p_2 p_3)^{2(1 - \epsilon)} I_{0+\epsilon \{-1+\epsilon, -1+\epsilon, -1+\epsilon\}}.
\end{equation}
Following the method described in section \eqref{sec:schemechangedetails}, we can now change the regularisation scheme from $-u = v_1 = v_2 = v_3$ to a general scheme with arbitrary $u$ and $v_j$, $j=1,2,3$. This concludes our calculation of the master integral leading to the final result as presented in \eqref{e:masterap}.

\section{Discussion}

In this paper we showed how to evaluate the integrals needed for the computation of momentum-space 3-point functions of operators of integer dimension in any CFT.  Together with the results in \cite{Bzowski:2013sza,Scalars}, one can now obtain explicit expressions for all such scalar 3-point functions and for all tensorial correlators that do not require renormalisation.  The results here are also sufficient  
for the computation of tensorial correlators that do require renormalisation, once this renormalisation  has been carried out as 
discussed in
\cite{Bzowski:2017poo, Bzowski:2018fql}. 

After reducing to triple-$K$ integrals, 
the most nontrivial cases are those with dimensions satisfying the triangle inequalities in \eqref{e:triangle} 
and we developed a comprehensive procedure for the evaluation of all such integrals.
We showed that all such integrals can be reduced  to the master integral $I_{0\{111\}}$ and we computed this integral, with the answer 
given in \eqref{e:masterap}.

In all remaining cases the computation is more straightforward. If the spacetime dimension is odd, then all Bessel $K$ functions appearing in the triple-$K$ integrals become elementary and can be computed straightforwardly (the result is given in appendix \ref{sec:halfint}).  If the spacetime dimension is even and the conditions above are not satisfied, the correlator can be extracted from the divergent part of the triple-$K$ integrals, as discussed in \cite{Scalars}.

\section*{Acknowledgements} 

AB and KS would like to thank the Galileo Galilei Institute in Florence for support and hospitality during the workshop ``Holographic Methods for Strongly Coupled Systems'' and
PM thanks the Centre du Recherches Mathematiques, Montreal. 
KS gratefully acknowledges support from the Simons Center for Geometry and Physics, Stony Brook University and the 
``Simons Summer Workshop 2015:  New advances in Conformal Field Theories'' during which some of the research for this paper was performed. 
AB is supported by the Interuniversity Attraction Poles Programme
initiated by the Belgian Science Policy (P7/37) and the European
Research Council grant no.~ERC-2013-CoG 616732 HoloQosmos.
AB would like to thank COST for partial support via the STMS grant COST-STSM-MP1210-29014. 
PM is supported by the STFC Consolidated Grant ST/L00044X/1. AB and PM would like to thank the University of Southampton for hospitality during parts of this work.

\appendix

\section{Useful formulae}

The Bessel function $I$, also known as the modified Bessel function of the first kind, is defined by the series
\begin{equation} \label{e:serI}
I_\nu(x) = \sum_{j=0}^\infty \frac{1}{j! \Gamma(\nu + j + 1)} \left( \frac{x}{2} \right)^{\nu + 2j}, \quad \nu \neq -1, -2, -3, \ldots
\end{equation}
The Bessel function $K$, or modified Bessel function of the second kind, is defined by
\begin{align}
K_\nu(x) & = \frac{\pi}{2 \sin (\nu \pi)} \left[ I_{-\nu}(x) - I_{\nu}(x) \right], \quad \nu \notin \Z, \label{e:defK} \\
K_n(x) & = \lim_{\epsilon \rightarrow 0} K_{n + \epsilon}(x), \quad n \in \Z.
\end{align}
For $x > 0$, the finite point-wise limit  exists for any integer $n$.

The series expansion of the Bessel function $K_\nu$ for $\nu \notin \Z$ is given directly in terms of the expansion \eqref{e:serI} via the definition \eqref{e:defK}. In particular
\begin{equation} \label{e:expK}
K_{\nu}(x) = \sum_{j=0}^\infty \left[ a^{-}_j(\nu) x^{-\nu + 2j} + a^{+}_j(\nu) x^{\nu + 2j} \right], \quad \nu \notin \Z,
\end{equation}
where the expansion coefficients read
\begin{equation}
a_j^{\sigma}(\nu) = \frac{(-1)^j \Gamma(- \sigma \nu - j)}{2^{\sigma \nu + 2 j + 1} j!}, \ \sigma \in \{ \pm 1 \}. \label{e:cf_ap}
\end{equation}
For non-negative integer index $n$, the expansion reads instead
\begin{align} \label{e:expKn}
K_n(x) & = \frac{1}{2} \left( \frac{x}{2} \right)^{-n} \sum_{j=0}^{n-1} \frac{(n-j-1)!}{j!} (-1)^j \left( \frac{x}{2} \right)^{2j} \nn\\
& \qquad + \: (-1)^{n+1} \log \left( \frac{x}{2} \right) I_n(x) \nn\\
& \qquad + \: (-1)^n \frac{1}{2} \left( \frac{x}{2} \right)^n \sum_{j=0}^\infty \frac{\psi(j+1) + \psi(n+j+1)}{j! (n+j)!} \left( \frac{x}{2} \right)^{2j},
\end{align}
where $\psi$ is the digamma function. At large $x$, the Bessel functions have the asymptotic expansions 
\begin{equation} \label{e:asymIK}
I_\nu(x) = \frac{1}{\sqrt{2 \pi}} \frac{e^x}{\sqrt{x}} + \ldots, \qquad K_\nu(x) = \sqrt{\frac{\pi}{2}} \frac{e^{-x}}{\sqrt{x}} + \ldots, \quad \nu \in \R.
\end{equation}

Appell's $F_4$ function  can be defined by the double series \cite{Appell,Erdelyi}
\begin{equation}
F_4(\alpha, \beta; \gamma, \gamma'; \xi, \eta) = \sum_{i,j = 0}^\infty \frac{(\alpha)_{i+j} (\beta)_{i+j}}{(\gamma)_i (\gamma')_j i! j!} \xi^i \eta^j, \quad \sqrt{|\xi|} + \sqrt{|\eta|} < 1,
\end{equation}
where $(\alpha)_i$ is a Pochhammer symbol. Notice that
\begin{equation}
F_4(\alpha, \beta; \gamma, \gamma'; \xi, \eta) = F_4(\beta, \alpha; \gamma, \gamma'; \xi, \eta) = F_4(\alpha, \beta; \gamma', \gamma; \eta, \xi).
\end{equation}
The series representation, however, is not very useful as in our case
\begin{equation}
\xi = \frac{p_1^2}{p_3^2}, \qquad \qquad \eta = \frac{p_2^2}{p_3^2}
\end{equation}
and the series only converges when $p_3 > p_1 + p_2$, which is opposite to the triangle inequality obeyed by physical momentum configurations.

The following reduction formulae can be found in \cite{Prudnikov} or \cite{Erdelyi}
\begin{align}
& F_4 \left( \alpha, \beta; \alpha, \beta; - \frac{x}{(1-x)(1-y)}, - \frac{y}{(1-x)(1-y)} \right)  \nn\\
& \qquad\qquad = \frac{(1-x)^\beta (1-y)^\alpha}{1-x y}, \label{e:redform1} \\
& F_4 \left( \alpha, \beta; \beta, \beta; - \frac{x}{(1-x)(1-y)}, - \frac{y}{(1-x)(1-y)} \right) \nn\\
& \qquad\qquad = (1-x)^\alpha (1-y)^\alpha {}_2 F_1(\alpha, 1+\alpha-\beta; \beta; x y), \\
& F_4 \left( \alpha, \beta; 1+\alpha-\beta, \beta; - \frac{x}{(1-x)(1-y)}, - \frac{y}{(1-x)(1-y)} \right) \nn\\
& \qquad\qquad = (1-y)^\alpha {}_2 F_1 \left(\alpha, \beta; 1+\alpha-\beta; - \frac{x(1-y)}{1-x} \right), \\
& {}_2 F_1 (2 \nu - 1, \nu; \nu; x) = (1 - x)^{1-2\nu}. \label{e:redform2}
\end{align}

\section{Triple-$K$ and momentum-space integrals} \label{sec:momentum}

Let $K_{d \{ \delta_1 \delta_2 \delta_3\}}$ denote a massless scalar 1-loop 3-point momentum-space integral,
\begin{equation}
K_{d \{ \delta_1 \delta_2 \delta_3\}} = \int \frac{\D^d \bs{k}}{(2 \pi)^d} \frac{1}{k^{2 \delta_3} | \bs{p}_1 - \bs{k} |^{2 \delta_2} | \bs{p}_2 + \bs{k} |^{2 \delta_1}}.
\end{equation}
Any such integral can be expressed in terms of triple-$K$ integrals and vice versa. For scalar integrals, the relation reads
\begin{equation}
K_{d \{ \delta_1 \delta_2 \delta_3\}} =\frac{2^{4 - \frac{3d}{2}}}{\pi^{\frac{d}{2}}}\times \frac{I_{\frac{d}{2} - 1 \{ \frac{d}{2} + \delta_1 - \delta_t, \frac{d}{2} + \delta_2 - \delta_t, \frac{d}{2} + \delta_3 - \delta_t \}}}{\Gamma(d - \delta_t) \Gamma(\delta_1) \Gamma(\delta_2) \Gamma(\delta_3)},
\end{equation}
where $\delta_t = \delta_1 + \delta_2 + \delta_3$. Its inverse reads
\begin{align}
I_{\alpha \{ \beta_1 \beta_2 \beta_3 \}} & = 2^{3 \alpha - 1} \pi^{\alpha + 1} \Gamma \left( \frac{\alpha + 1 + \beta_t}{2} \right) \prod_{j=1}^3 \Gamma \left( \frac{\alpha + 1 + 2 \beta_j - \beta_t}{2} \right)  \nn\\
& \qquad\qquad \times K_{2 + 2 \alpha, \{ \frac{1}{2} (\alpha + 1 + 2 \beta_1 - \beta_t), \frac{1}{2} (\alpha + 1 + 2 \beta_2 - \beta_t), \frac{1}{2} (\alpha + 1 + 2 \beta_3 - \beta_t) \}},
\end{align}
where $\beta_t = \beta_1 + \beta_2 + \beta_3$.

All tensorial massless 1-loop 3-point momentum-space integrals can be also expressed in terms of a number of triple-$K$ integrals when their tensorial structure is resolved by standard methods, \textit{e.g.}, \cite{Passarino:1978jh,Denner:2005nn}. For exact expressions in this case see appendix A.3 of \cite{Bzowski:2013sza}.

\section{Half-integral betas}
\label{sec:halfint}

Bessel $K$ functions with half-integral indices reduce to elementary functions meaning the corresponding triple-$K$ integrals can be evaluated with ease.  We present below a complete expression valid for any triple-$K$ integral in which all $\beta_j$ are positive half-integral numbers.

A Bessel function $K$ with a half-integral index is equal to
\begin{equation}
K_{\beta}(x) = \frac{e^{-x}}{\sqrt{x}} \sum_{j=0}^{|\beta| - \frac{1}{2}} \frac{c_j(\beta)}{x^j}, \ \beta \in \Z + \frac{1}{2},
\end{equation}
where the coefficients are
\begin{equation} \label{e:cfa}
c_j(\beta) = \sqrt{\frac{\pi}{2}} \frac{ \left( |\beta| - \frac{1}{2} + j \right)!}{2^j j! \left(|\beta| - \frac{1}{2} - j \right)!}.
\end{equation}
Any triple-$K$ integral for which all $\beta_j$ are half-integer then evaluates to
\begin{align} \label{e:Ihalf}
I_{\alpha \{ \beta_1 \beta_2 \beta_3 \}} & = \sum_{k_1 = 0}^{|\beta_1| - \frac{1}{2}} \sum_{k_2 = 0}^{|\beta_2| - \frac{1}{2}} \sum_{k_3 = 0}^{|\beta_3| - \frac{1}{2}} \Gamma \left( \alpha - \frac{1}{2} - k_t \right) p_t^{\frac{1}{2} + k_t - \alpha}  \nn\\
& \qquad\qquad \times p_1^{\beta_1 - k_1 - \frac{1}{2}} p_2^{\beta_2 - k_2 - \frac{1}{2}} p_3^{\beta_3 - k_3 - \frac{1}{2}} c_{k_1}(\beta_1) c_{k_2}(\beta_2) c_{k_3}(\beta_3),
\end{align}
where $k_t = k_1 + k_2 + k_3$ and $p_t = p_1 + p_2 + p_3$, and the value of $\alpha$ is arbitrary. 

For 3-point functions featuring operators of integral dimensions in \emph{odd}-dimensional spacetimes $\alpha$ is half-integer. In such cases the gamma function in the expression above may become singular, provided the condition \eqref{e:cond1} is satisfied.
Assuming $\alpha$ is half-integer, one can regulate \eqref{e:Ihalf} in a scheme with all $v_j = 0$ by shifting $\alpha \mapsto \reg{\alpha} = \alpha + u \epsilon$. The usual expansion of the gamma function can then be applied. 
To change the regularisation scheme to one with non-vanishing $v_j$ we then follow the procedure of section \ref{sec:schemechangedetails}.

\section{Derivation of scheme-changing formula}
\label{sec:proof}

Our goal in this section is to derive the formula \eqref{e:tobe_change} for changing the regularisation scheme.  Following the discussion in section \ref{sec:schemechangedetails},
our first move is to split the regulated triple-$K$ integral into three parts,
\begin{equation}
I_{\reg{\alpha} \{ \reg{\beta}_j \}} = I^{\text{(div)}}_{\reg{\alpha} \{ \reg{\beta}_j \}} + I^{\text{(lower)}}_{\reg{\alpha} \{ \reg{\beta}_j \}} + I^{\text{(upper)}}_{\reg{\alpha} \{ \reg{\beta}_j \}},
\end{equation}
where $I^{\text{(div)}}$ is given by \eqref{e:div} and
\begin{align}
I^{\text{(upper)}}_{\reg{\alpha} \{ \reg{\beta}_j \}} & = \int_{\mu^{-1}}^{\infty} \D x \: x^{\reg{\a}} \prod_{j=1}^3 p_j^{\reg{\b}_j} K_{\reg{\b}_j}(p_j x), \\
I^{\text{(lower)}}_{\reg{\alpha} \{ \reg{\beta}_j \}} & = \int_0^{\mu^{-1}} \D x \: x^{\reg{\a}} \prod_{j=1}^3 p_j^{\reg{\b}_j} K_{\reg{\b}_j}(p_j x) - I^{\text{(div)}}_{\reg{\alpha} \{ \reg{\beta}_j \}}. \label{e:idown}
\end{align}
Both $I^{\text{(upper)}}$ and $I^{\text{(lower)}}$ as defined here are clearly finite in the limit $\epsilon \rightarrow 0$.
$I^{\text{(upper)}}$ converges for any $\reg{\alpha}$ and $\reg{\beta}$ provided $\mu^{-1}>0$ and $I^{\text{(lower)}}$ is finite since all divergent terms have been explicitly subtracted in its definition.

We now want to show that the finite pieces ({\it i.e.,} terms of order $\ep^0$) in $I^{\text{(upper)}}$ and $I^{\text{(lower)}}$ are independent of $u$ and $v_j$.  The finite part of the difference between a regulated triple-$K$ integral and its divergent part $I^{\text{(div)}}$ is then scheme independent, reproducing \eqref{e:schemeindep} and \eqref{e:tobe_change}.

To show the finite part of $I^{\text{(upper)}}$ is scheme independent, note  that
\begin{align}
\lim_{\epsilon \rightarrow 0} I^{\text{(upper)}}_{\reg{\alpha} \{ \reg{\beta}_j \}} & = \int_{\mu^{-1}}^{\infty} \D x \: \lim_{\epsilon \rightarrow 0} x^{\reg{\a}} \prod_{j=1}^3 p_j^{\reg{\b}_j} K_{\reg{\b}_j}(p_j x) 
= \int_{\mu^{-1}}^{\infty} \D x \: x^{\a} \prod_{j=1}^3 p_j^{\b_j} K_{\b_j}(p_j x).
\end{align}
For any $\mu^{-1}>0$, the exchange of the integral and the limit here is justified by the dominated convergence theorem.  The finite part of $I^{\text{(upper)}}$ is therefore scheme independent since the right-most expression is independent of $u$ and $v_j$.

Similar arguments can be used to show that  the finite part of $I^{\text{(lower)}}$ is scheme independent. First, we series expand all Bessel functions in the integrand of the triple-$K$ integral.  In the following, it will be useful to denote the coefficient of $x^{A + B \epsilon} \log^N x$ in this expansion as $c_{A + B \epsilon, N}$. 
Note that the value of $A$ is independent of $u$ and $v_j$, whereas $B=B(u,v_j)$ is scheme dependent.
We can then write
\begin{equation}
I^{\text{(lower)}}_{\reg{\alpha} \{ \reg{\beta}_j \}} = \sum_{\substack{A, \: B \in \R \\ A \neq -1 \\ N \in \{0,1,\ldots\}}} \int_0^{\mu^{-1}} \D x \: x^{A + B(u,v_j) \epsilon} \log^N x \: c_{A + B(u,v_j) \epsilon, N}, \label{e:idownfin}
\end{equation}
where the sum is taken over all terms appearing in the series expansion of the triple-$K$ integrand except for those of the form $x^{-1 + O(\epsilon)}$ ({\it i.e.,} for which $A=-1$).  Terms of this form appeared in $I^{\text{(div)}}$ but were subtracted in the definition  of $I^{\text{(lower)}}$ (see \eqref{e:idown}). 
Our exchange in the order of summation and integration going from \eqref{e:idown} to \eqref{e:idownfin} is justified by Fubini's theorem, noting that the sum converges absolutely.
The natural analytic continuation of \eqref{e:analcont} can then be used to extend the result to any values of $\alpha$ and $\beta_j$.

For purposes of illustration let us now concentrate on two special cases; the general case can then be handled by similar arguments.
As a first case, consider the regularisation scheme where all  $v_j = 0$.  From the series expansion  \eqref{e:expKn}, all $c_{A + B \epsilon, N}$ in \eqref{e:idownfin} are then finite and scheme-independent. The remaining integral is given by \eqref{e:analcontexp}, and since $A \neq -1$, the finite part of $I^{\text{(lower)}}$ is indeed scheme-independent as we wished to show. 

As our second case, consider the opposite situation where all $v_j \neq 0$. The $c_{A + B \epsilon, N}$ then vanish for all $N > 0$ while $c_{A + B \epsilon, 0}$ can be expanded in terms of the Bessel expansion coefficients $a_j^\sigma(\beta)$ in \eqref{e:cf}. The result takes a form similar to \eqref{e:div1}, namely
\begin{equation} \label{e:findown1}
I^{\text{(lower)}}_{\reg{\alpha} \{ \reg{\beta}_j \}} = \sum_{\text{cond}}\!{}'\:\frac{\mu^{-W}}{W} \prod_{j=1}^3 p_j^{(1 + \sigma_1) \reg{\beta}_j + 2 n_j} a_{n_j}^{\sigma_j}(\reg{\beta}_j),
\end{equation}
where
\begin{equation} \label{e:W}
W = \reg{\alpha} + 1 + \sum_{j=1}^3 ( \sigma_j \reg{\beta}_j + 2 n_j ).
\end{equation}
This time, however, the summation runs over all terms in the complement of those present in \eqref{e:div1}, which we have indicated with a prime on the summation sign.  In other words, the sum runs over all $\sigma_1, \sigma_2, \sigma_3 \in \{ \pm 1 \}$ and non-negative integers $n_1, n_2, n_3$ such that the condition \eqref{e:cond3} is \emph{not} satisfied.

Clearly $1/W$ has a finite and scheme-independent limit as $\ep\rightarrow 0$ given that  \eqref{e:cond3} is not satisfied. Moreover, if a given $\b_j$ is non-integer, $a_{n_j}^{\sigma_j}(\reg{\b}_j)$ has a finite and scheme-independent limit as $\epsilon \rightarrow 0$ as follows from its definition \eqref{e:cf}. If $\b_j \in \Z$, on the other hand, $a_{n_j}^{\sigma_j}(\reg{\b}_j)$ may diverge as $\epsilon \rightarrow 0$. More precisely, $a^{-}_{n_j}(\reg{\b}_j)$ has a finite, scheme-independent limit if $n_j < \b_j$ but otherwise diverges, while $a^{+}_{n_j}(\reg{\b}_j)$ always diverges for integer $\b_j$. 
Nevertheless, such divergences do not lead to a divergence in $I^{\text{(lower)}}$
since the divergent contributions from 
 $a^{+}_{n_j}(\reg{\b}_j)$ and $a^{-}_{n_j + \b_j}(\reg{\b}_j)$ cancel.
To see this,  consider for concreteness the case $\beta_1 \in \Z$.  Denoting the values of \eqref{e:W} with given $n_1$ and $\sigma_1 = \pm 1$ (but all other parameters fixed) by $W^{\pm}_{n_1}$, the corresponding contributions to 
\eqref{e:findown1} take the form
\begin{equation}
\frac{\mu^{-W^{+}_{n_1}}}{W^{+}_{n_1}} p_1^{2 \reg{\b}_1 + 2 n_1} a_{n_1}^{+}(\reg{\b}_1) + \frac{\mu^{-W^{-}_{n_1+\b_1}}}{W^{-}_{n_1 + \b_1}} p_1^{2 (\b_1 + n_1)} a_{n_1 + \b_1}^{-}(\reg{\b}_1).
\end{equation}
By inspection, however, this expression has a finite and scheme-independent limit as $\epsilon \rightarrow 0$. Since every $a^{+}_{n_1}(\reg{\b}_1)$ can be matched with its corresponding $a^{-}_{n_1 + \b_1}(\reg{\b}_1)$, the expression \eqref{e:findown1} therefore has a finite and scheme-independent limit.  The finite part of $I^{\text{(lower)}}$ is thus again scheme-independent as we wished to show.

\bibliography{reduction}
\bibliographystyle{JHEP} 

\end{document}